\documentclass{aa}
\usepackage{txfonts}
\usepackage{graphicx}
\usepackage[comma]{natbib}
\usepackage{rotating}
\usepackage{aalongtable}
\bibpunct{(}{)}{;}{a}{}{,}
\newcommand{\be}{\begin{equation}}
\newcommand{\ee}{\end{equation}}
\newcommand{\lb}[1]{\label{#1}}

\newcommand{\ssty}{\scriptscriptstyle}
\newcommand{\tsty}{\textstyle}
\newcommand{\lcdm}[1]{#1_{\ssty \Lambda}}
\newcommand{\void}[1]{#1_{{\tsty {\ssty V}}}}
\newcommand{\biglcdm}[1]{\left. #1 \right|_{\ssty \Lambda}}
\newcommand{\bigvoid}[1]{\left. #1 \right|_{{\tsty {\ssty V}}}}

\newcommand{\etal}{et al.\ }

\newcommand{\dl}{d_{\ssty L}}
\newcommand{\da}{d_{\ssty A}}
\newcommand{\dg}{d_{\ssty G}}

\newcommand{\dc}{r}

\vspace{+1cm}
\begin{document}
\title{Cosmological model dependence of the galaxy luminosity function:
far-infrared results in the Lema\^{\i}tre-Tolman-Bondi model}
\titlerunning{Void-cosmology systematics on the FIR LF}
\author{A.~Iribarrem \inst{\ref{OV},\ref{ESO}} \thanks{iribarrem@astro.ufrj.br}
\and P.~Andreani \inst{\ref{ESO}}
\and C. ~Gruppioni \inst{\ref{INAF}}
\and S. ~February \inst{\ref{UCT}}
\and M.~B.~Ribeiro \inst{\ref{IF}}
\and S. ~Berta \inst{\ref{MPE}}
\and E. ~Le Floc'h \inst{\ref{CEA}}
\and B. ~Magnelli \inst{\ref{MPE}}
\and R. ~Nordon \inst{\ref{MPE}}
\and P. ~Popesso \inst{\ref{MPE}}
\and F. ~Pozzi \inst{\ref{UBol}}
\and L. ~Riguccini \inst{\ref{CEA}}
\authorrunning{Iribarrem, Andreani, Gruppioni \etal}
}
\institute{
Observat\'orio do Valongo, Universidade Federal do Rio de Janeiro,
Ladeira Pedro Antonio 43, 20080-090, Rio de Janeiro, Brazil.
\label{OV}
\and
European Southern Observatory (ESO),
Karl-Schwarzschild-Stra\ss e 2, 85748 Garching, Germany.
\label{ESO}
\and
INAF -- Osservatorio Astronomico di Bologna,
via Ranzani 1, I-40127, Bologna, Italy.
\label{INAF}
\and
Astrophysics, Cosmology and Gravitation Centre,
and Department of Mathematics and Applied Mathematics,
University of Cape Town, Rondebosch 7701, Cape Town, South Africa.
\label{UCT}
\and
Instituto de F\'\i sica, Universidade Federal do Rio de Janeiro,
CP 68532, 21941-972, Rio de Janeiro, Brazil.
\label{IF}
\and
Max-Planck-Institut f\"{u}r Extraterrestrische Physik (MPE),
Postfach 1312, D-85741, Garching, Germany.
\label{MPE}
\and
CEA-Saclay, Service d'Astrophysique,
F-91191, Gif-sur-Yvette, France.
\label{CEA}
\and
Dipartimento di Astronomia, Universit\`{a} di Bologna,
via Ranzani 1, I-40127, Bologna, Italy.
\label{UBol}
}
\date{ }
\abstract{}
{This is the first paper of a series aiming at investigating galaxy formation and evolution in the { giant-void class of the Lema\^\i tre-Tolman-Bondi
(LTB) models that best fits current cosmological observations}.
Here we investigate {the Luminosity Function (LF) methodology, and how} its
estimates would be affected by a change on the cosmological model assumed in its computation.
Are the current observational constraints on the allowed Cosmology enough
to yield robust LF results?}
{We use the far-infrared source catalogues built on the observations performed with the
Herschel/PACS instrument, and selected as part of the PACS evolutionary
probe (PEP) survey. Schechter profiles are obtained in redshift bins up
to z $\approx$ 4, assuming comoving volumes in both the standard model,
that is, Friedmann-Lema\^\i tre-Robertson-Walker metric with a perfect fluid
energy-momentum tensor, and non-homogeneous LTB
dust models, parametrized to fit the current combination of results
stemming from the observations of supernovae Ia, the cosmic microwave
background, and baryonic acoustic oscillations.}
{We find that the luminosity functions computed assuming both the standard model
and LTB void models show in general good agreement. However, the
faint-end slope in the void models shows a significant departure from the
standard model up to redshift 0.4. We demonstrate that this result is not artificially caused by the used LF estimator
which turns out to be robust under the differences in matter-energy density profiles of the models.}
{The differences found in the LF slopes at the faint end are due to variation in the luminosities 
of the sources, which depend on the geometrical part of the model. It follows that either the
standard model is over-estimating the number density of faint sources or the void models are
under-estimating it.}

\maketitle

\section{Introduction}

The luminosity function (LF) is an important observational tool for galaxy evolution
studies, as it encodes the observed distribution of galaxies in volumes and luminosities.
However, a cosmological model must be assumed in its estimation, rendering it model
dependent. { On the other hand, the precision of the current constraints on
the cosmological model might arguably be enough to yield an LF that has significantly the same shape in all models allowed by the observations. To investigate this assertion, it is necessary to compute the LF considering one
such alternative model, and perform a statistical comparison with the LF obtained
assuming the standard model.}

The currently favoured theory for explaining the shape and redshift evolution
of the LF is that the dark matter haloes grow up hierarchically by merging, and that
baryonic matter trapped by those haloes condense to form galaxies. { Gastrophysics
processes (gas cooling, high redshift photoionization, feedbacks), are then responsible
to reproduce the shape of the luminosity function of galaxies starting from the
Dark Matter halo mass function \citep{2003ApJ...599...38B}. The usual approaches}
in the context of the standard model of cosmology, { are either to use}
Semi-Analytical Models to parameterize such processes, \citep[e.g.][]{2010MNRAS.405.2717N},
or empirical { models,  \citep[e.g.][]{2003MNRAS.339.1057Y,2009MNRAS.392.1080S,2011ApJ...736...59Z},
to allocate galaxies as a function of halo mass, both} built on top of a
dark matter hierarchical merger tree created by simulations,
like the Millennium simulation, \citep{2005Natur.435..629S,2009MNRAS.398.1150B}.

{ It is well-established by observations made at many different wavelengths
\citep[some recent examples include][]{2010A&A...523A..74V,2011AJ....142...41R,2012ApJ...748...10C, 2012MNRAS.421.3060S,2013MNRAS.428..291P,2013MNRAS.429..881S},
and particularly in the IR \citep{2006MNRAS.370.1159B,2007ApJ...660...97C, 2010A&A...515A...8R,2011A&A...528A..35M,2013MNRAS.429.1113H},
that the LF shows significant evolution with the redshift.}
In practice the LF is traditionally computed using the comoving volume, which does not
stem directly from the observations, but is rather derived from it assuming a
cosmological model, with a well-defined metric that translates redshifts
into distances. The effects of the expanding space-like hipersurfaces can, therefore, be
successfully factored out of the observations, {\em up to the limits where the assumed
cosmological model holds}. The last remark is of special importance, since
\citep{1997MNRAS.292..817M} proved that any spherically symmetric set of observations,
{ like redshift surveys},
can be fitted simply by spatial non-homogeneities in a more general cosmological model that
assumes a Lema\^\i tre-Tolman-Bondi (LTB) line element and a dust-like energy-momentum
tensor, {\em regardless of any evolution of the sources}. { As a consequence, the
reported redshift evolution of the LF could, in principle, be caused by a non-homogeneity
on the cosmology, at the scale of the observations.} It is therefore crucial for
galaxy evolution theories, and past-lightcone studies such as \citep{2003ApJ...592....1R,
2007ApJ...657..760A, 2008A&A...488...55R, 2012A&A...539A.112I, 2012ApJ...752..113H,
2012MNRAS.tmp.3286D}, 
that the underlying cosmological model be well-established by independent observations.

Results from many independent cosmological observations fit together
in a coherent picture under the $\Lambda$CDM model, \citep[e.g.][]{2009ApJS..180..330K},
this being the main source of its present success as it is nowadays adopted as the
main cosmological model.

One of the observational results, arguably key in selecting the $\Lambda$CDM parametrization
for a Friedmann-Lema\^\i tre-Robertson-Walker (FLRW) perfect fluid model, is the
dimming in the redshift-distance relation of supernovae Ia, first obtained independently
by \citep{1998AJ....116.1009R} and \citep{1999ApJ...517..565P}. This has led to the
re-introduction of the cosmological constant $\Lambda$ in Einstein's field equations, and
the further interpretation of it as an exotic fluid, {\em dark energy}, accelerating the
expansion of the Universe.


{ Despite the many empirical successes of the standard model, understanding of the
physical nature of dark energy is still lacking. This fact has encouraged many authors
to investigate viable alternatives to it}, like modified gravity \citep{2010LNP...800...99T},
the effect of small-scale spatial non-homogeneities of the matter content in the estimation of the
cosmological model parameters \citep{2012MNRAS.tmpL.504B}, often called the
backreaction effect on cosmology, \citep{2011arXiv1109.2314C, 2011CQGra..28p4010C,
2012PhRvD..86d3506C, 2012A&A...538A.147W}, or non-homogeneous cosmological models
\citep{2007astro.ph..2416C, 2011CQGra..28p4002B, 2011CQGra..28p4001E}.

Many recent works have advanced our understanding of non-homogeneities and, particularly,
of LTB models. From practical issues
like those related to possible dimming, or brightening, of point-like sources due to the
narrowness of their observed beams, as compared to the typical smoothing scales in
standard model simulations \citep{2012MNRAS.426.1121C}, or the possibility of
accounting for the anomalous primordial Lithium abundances \citep{2012GReGr..44..567R},
passing through development of the models themselves, like in
\citep{2009PhRvD..79d3501H, 2010GReGr..42.1935A, 2012MNRAS.419.1937M, 2012GReGr..44.3197H,
2012PhRvD..85j3511N, 2012PhRvD..85j3512B, 2012arXiv1209.4078V, 2012ApJ...748..111W,
2012JCAP...01..043H},
to several tests and fits to different observations, like those in
\citep{2010MNRAS.405.2231F, 2011JCAP...09..011B, 2013ApJ...762L...9H, 2012PhRvD..85b4002B,
2012arXiv1208.4534D}, much have been done to establish non-homogeneity as a
well-grounded modification of the standard cosmology. Despite the recent interest in these
kind of models, to date no work has aimed at studying galaxy evolution on non-homogeneous cosmologies.

By restricting the available models to those which are well constrained by a wealth
of observations, we focus on the question:
given that the current observations still allow a certain degree of freedom
for the cosmological model, are these constraining enough to yield a robust LF estimation,
or are our statistical conclusions still dependent on the model? And how?

To address this question, one needs at least two different cosmological models, in a
sense of a set of equations that are a solution for the Einstein's field equations,
both parametrized to fit the whole set of available observations. Therefore, for the purpose
stated above, the parametrization of \citep{2008JCAP...04..003G} for the LTB dust model is sufficient.



We start from the far-infrared (FIR) LF which has been recently well established by
\citep{2013MNRAS.tmp.1158G}, { using combined data obtained on the PACS \citep{2010A&A...518L...2P},
and SPIRE \citep{2010A&A...518L...3G} instruments aboard the Herschel \citep{2010A&A...518L...1P} 
space telescope, as part of the PEP, {\em the {\em PACS} Evolutionary Probe} \citep{2011A&A...532A..90L},
and HerMES, {\em the {\em Herschel} Multi-tiered Extragalactic Survey} \citep{2012MNRAS.424.1614O} surveys.}
We use this sample because of its huge wealth of observations spanning
from UV to the far-IR and it is the most complete one in terms of wavelength coverage.
In future works, we intend to investigate the effect on LF
when changing the underlying cosmology as a function of wavelength.
In fact, the depth of the survey, or the relative
depths at different wavelengths may also play a role.

\citep{2013MNRAS.tmp.1158G} have used the PEP datasets to derive evolutionary properties
of far-IR sources in the standard cosmology. We aim at using the same catalogues and methodology used
by \citep{2013MNRAS.tmp.1158G} to assess them in alternative cosmologies.
 We compute the rest-frame
monochromatic 100 $\mu$m and 160 $\mu$m, together with the total IR LFs in the
GBH void models described in \citep{2012JCAP...10..009Z}. We then compare the
redshift evolution of the luminosity functions in both standard and void models.

Although the present work uses both the standard and alternative cosmological models, it
does not aim at model selection, that is, making a comparison of the models {\em themselves}.
It is common to assume that works which deal with alternative cosmologies always have the goal
of testing the models directly. However, this is not always the case.

{\em This work is not about testing alternative cosmologies}. Since the beginning of Observational
Cosmology it has been clear that testing a cosmological model using galaxy surveys is extremely
difficult. That is because of the degeneracy between the intrinsic evolution of the sources and the
relativistic effects caused by the underlying cosmology: our understanding of galaxies is still
far from allowing us to treat them as standard candles. Besides, the data used in this work comes
from a survey not nearly wide enough to compute meaningful angular correlation functions.
The luminosity functions computed here depend, by definition, on the cosmological model assumed
in its computation, and therefore cannot yield any independent conclusion about which model is
the best fit. This is not the goal here.

Instead, by acknowledging the fact that the computation of the luminosity functions depend
on the cosmological model, we aim to assess how robust the luminosity function results are
if the effective constraints on the cosmological model, like the Hubble diagram of a survey of
standard candles like the SNe Ia, or the power spectrum of the Cosmic Microwave Background,
are imposed.
In other words, this work's main interest is galaxy evolution models, and their possible
dependency on the cosmological model but, not the cosmological models themselves.

A couple of recent papers have noteworthy similarities with the present work. \citep{2012ApJ...754..131K}
uses the near-infrared luminosity function of galaxies in the 0.1 $< z <$ 0.3 redshift range
to probe the central underdensity predicted by the void models. By assuming the standard
model line element, the authors argue that the presence of a local underdensity would lead
to an over-estimation of the normalization of their LFs. \citep{2012MNRAS.426.2566M}
discuss the effect of the cosmology dependence of the distance-redshift relation on the
clustering of galaxies. Apart having different goals, as stated above, the present work
differs from the ones above in that it assumes the LTB void models in all the steps
of the computation of its results.

The paper is divided as follows: in \S\ref{methods} we describe the dataset extracted
from the PEP multi-wavelength catalogs, and discuss the method used for
the estimation of the LFs in both cosmologies. In \S \ref{LTB} we
briefly describe the parameterization of the void models used, and obtain
expressions for the luminosity and the comoving distances in those models.
In \S \ref{results} we present the LF computed in both cosmological models, as well
as analytical fits to them. In \S \ref{discussion} we perform quantitative comparisons
of the LFs and their evolution in the different models.
We present our conclusions in \S \ref{conclusions}.

\section{luminosity functions}
\label{methods}

In this section we present and discuss the main results and equations in deriving
the LFs that are susceptible to a change if the underlying cosmology is modified.

\subsection{The PEP multi-wavelength samples}
\lb{samples}
We start from the multi-wavelength catalogues described in \citep{2011A&A...532A..49B}.
The sources in those catalogues were blind selected { in the following fields (effective
areas): GOODS-N (300 arcmin$^2$), GOODS-S (300 arcmin$^2$), COSMOS (2.04 deg$^2$),
and  ECDF-S (700 arcmin$^2$), as part} of the PEP survey, in the 100 and 160 $\mu$m filters
of Herschel/PACS.
{ The number of sources detected and the 3-$\sigma$ flux limits of this dataset are,
in the 100 and 160 $\mu$m passbands, respectively: 291 sources down to 3.0 mJy, and 
316 sources down to 5.7 mJy for GOODS-N; 717 down to 1.2 mJy, and 867 sources down to 2.4
mJy for GOODS-S; 5355 sources down to 5.0 mJy, and 5105 sources down to 10.2 mJy for COSMOS;
and finally, 813 sources down to 4.5 mJy, and 688 sources down to 8.5 mJy for ECDF-S.}
For each of those fields, in both bands individually, incompleteness corrections
for the number counts were computed by the authors using simulations.

The semi-empirical Spectral Energy Distribution (SED) models of \citep{2010A&A...518L..27G},
expanding on the ones of \citep{2007ApJ...663...81P}, were used to fit the photometry of the
objects using the LePhare code \citep{1999MNRAS.310..540A,2006A&A...457..841I}.
The code outputs, for each successfully fit source, a file with synthetic AB magnitudes,
$m_{\nu}$, in the wavelength range of the combined optical/NIR + FIR models. From that,
we compute the spectral density of flux,
$f(\nu)$ as,
\begin{equation}
\lb{ftomag}
f(\nu) = 10^{(23.9-m_{\nu})/2.5}.
\end{equation}

Sources without a redshift determination have been removed from the catalogues,
{ but no further redshift-based selection rule was applied. In the GOODS-N redshift
completeness is 100\% within ACS \citep{2010A&A...518L..30B} area, with 70\% of the redshifts there being
spectroscopic. These figures are 100(80)\% for the GOODS-S, within the MUSIC \citep{2006A&A...449..951G,
2012A&A...540A.109S} area; 93(40)\% for the COSMOS; and 88(25)\% for the ECDF-S fields}.
Non-detections in the 100 and 160 $\mu$m filters { were also removed}. Our final combined
samples have 5039 sources in the 100 $\mu$m band { (183 in the GOODS-N, 468 in the GOODS-S,
3817 in the COSMOS, and 578 in the ECDF-S fields);} and 5074 sources in the 160 $\mu$m one
{ (197 in the GOODS-N, 492 in the GOODS-S, 3849 in the COSMOS, and 547 in the ECDF-S fields).
Approximately 40\% of these sources were best fit by typical spiral SED templates,
7\% of those were best fit by starburst ones, another 7\% were found to be luminous IR galaxies,
and 46\% were best fit by obscured, or low-luminosity AGN templates.}
For a more complete description of the dataset used in this work, see \citep{2013MNRAS.tmp.1158G}.

We compute the rest frame total IR luminosity of each
source using its best fit SED, $f(\nu)$, by means of,
\begin{equation}
L_{\ssty IR} = 4 \pi \, (1+z) \, \dl (z)^2 \, \int_{8 \mu m}^{1000 \mu m} \,
f(\nu) d\nu \, .
\end{equation}

While, the rest-frame luminosity related to the observed flux $f_{\ssty R}$
at a given band $R$ can be obtained with,
\begin{equation}
\lb{rflum}
L_{\ssty R} = 4 \pi \, \nu_{\ssty R} \,  k_{\ssty R} \,
f_{\ssty R} \, \dl (z)^2,
\end{equation}
where $\dl$ is the luminosity distance in a particular cosmological model,
$\nu_{\ssty R}$ is the filter's effective frequency, at the observer's frame
(corresponding to wavelengths approximately 100 and 160 $\mu$m for the PACS bands
considered in the present work), and $k_{\ssty R}$ is the k-correction between
the observed frame flux $f_{\ssty R}$ in the $R$ band and its rest-frame flux,
at redshift $z$.

{
Because this paper deals with
more than one underlying metric, it is important to note that, even though the
relation between the cosmological redshift and the cosmological distances
depend on the metric -- thus affecting, for example, maximum redshift
estimates -- the redshift itself, and its effect on the SED of the sources, is
directly measurable. Therefore, even though the rest-frame luminosities
themselves depend on the cosmological model, the k-correction values depend
only on the redshift measurements. It is when translating the measured redshift
to an actual distance that a metric for the underlying spacetime is needed.

\subsection{k-corrections}
\lb{kcorrsec}
In the following discussion, all quantities are written in
frequency units. Primes are used to mark quantities evaluated at the
{\em source's rest frame}. We follow closely the derivation in
\citep{2002astro.ph.10394H}, but write the resulting k-correction in
terms of fluxes, instead of magnitudes.

The effect of the expansion of the metric over
the frequencies of the light arriving from each source is,
\begin{equation}
\lb{zfuncdef}
\nu' = (1+z) \nu,
\end{equation}
where $\nu'$ are rest-frame frequencies, measured by an observer in a
comoving frame with the source, $\nu$ are observed frequencies, measured
by an observer that is receding in relation to the source at a redshift $z$.

Fitting a SED template to the measured photometry for a given source, yields 
a model for its observed spectral density of flux, $f(\nu)$, over a range of
observed-frame frequencies. With that in hand, one can then compute the spectral
density of flux $f_{\ssty R}$, as measured by a given filter $R(\nu)$, in the
observed-frame, by means of the dimensionless convolution
(to ensure $f(\nu)$ and $f_{\ssty R}$ are both written in the same units),
\begin{equation}
\lb{fconv}
f_{\ssty R} = \int f(\nu) \, R(\nu) \, \frac{d\nu}{\nu}.
\end{equation}

To correctly account for the expansion effects when computing the rest-frame
spectral density of flux, $f'_{\ssty R}$, on the same passband $R$, one must first
redshift the filter function $R(\nu)$ $f(\nu)$ in the observed-frame, back to
source's rest-frame frequencies $f(\nu')$. Given the source's measured redshift $z$,
this can be done by means of equation \ref{zfuncdef}, yielding $R(\nu') = R[(1+z)\nu]$.
The rest-frame $R(\nu')$, can then be convolved with the observed-frame $f(\nu)$,
to yield the spectral density of flux, as measured by the passband $R$ at the source's
rest-frame, as,
\begin{equation}
\lb{zconv}
f'_{\ssty R} = \int f(\nu) \, R[(1+z)\nu] \, \frac{d\nu}{\nu}.
\end{equation}
Once $f'_{\ssty R}$ is obtained, the k-correction expressed in terms of densities of fluxes
is then,
\begin{equation}
\lb{kcorr}
k_{\ssty R} = \frac{f_{\ssty R}}{f'_{\ssty R}}.
\end{equation}
We note that a similar expression is used by \citep{2007AJ....133..734B}, based on the
derivation for the k-correction expressed in terms of magnitudes given in \citep{2002astro.ph.10394H}.

}
Next, we describe the use of the $1/V_{max}$ estimator, \citep{1968ApJ...151..393S},
in the computation of the LF of the samples.

\subsection{$1/V_{max}$ estimator}
\lb{vmaxlfsec}
The $1/V_{max}$ { \citep{1968ApJ...151..393S, 2011A&ARv..19...41J}} estimator for the LF
has the advantage of not assuming a parametric form in
its calculation. It also yields directly the comoving number density normalisation.
Recent results from \citep{2012MNRAS.426..531S} show that large-scale density variations can
introduce systematic errors in the subsequent parameters fitting.
Since we are dealing with different cosmological models that predict different density parameter
evolutions, it is important to check how dependent the method itself is on the cosmology.
We report in Appendix \ref{mock} how we built mock catalogues to check the effects of density variation,
similar to what is done in \citep{2000ApJS..129....1T}, and check that this
methodology is adequate for the purpose of the paper.

To compute the LF values using this method, we start by dividing each sample in redshift intervals,
$\Delta z$, with centre values $\bar{z}$, and in luminosity bins, $\Delta L$, with centre
values $\bar{L}$.
For each source in each ($\bar{z}$,$\bar{L}$) bin, we compute the maximum redshift
at which it would still be included in the survey. Given the corresponding flux limit for the field
where the source was detected, $f_{\ssty R, lim}$ \citep{2011A&A...532A..49B}, its measured
flux at that filter $R$, $f_{\ssty R}$, and its redshift, $z$, the highest redshift at which that source would
still be included, $\zeta$, can be obtained by means of the following relation\footnote{Since we are dealing with two observed-frame quantities,
there is no need to include any k-corrections in the equation~(\ref{eq:kcorr}).},
\begin{equation}
\lb{zmaxdef}
f_{\ssty R, lim} = \left[ \frac{\dl(z)}{\dl(\zeta)} \right]^2 \, f_{\ssty R}.
\label{eq:kcorr}
\end{equation}

If the maximum redshift for a given source is outside the redshift interval it
originally belongs, we use the upper limit of such interval, $z_{\ssty h}$, as the maximum redshift
instead. That is,
\begin{equation}
\lb{zmax}
z_{\ssty max} = min(z_{\ssty h},\zeta).
\end{equation}
The maximum comoving volume, $V_{\ssty max}$, enclosing each source is then,
\begin{equation}
\lb{Vc}
V_{\ssty max} = \sum_{\ssty k} \frac{S_{\ssty k}}{3} \int_{z_{\ssty l}}^{z_{\ssty max}}
w_{\ssty k}(z) \, r(z)^2 \frac{dr}{dz} dz,
\end{equation}
where the sum is over the $k$ fields where the source would have been included, $S_{\ssty k}$
is the area of the field where the source was detected, $z_{\ssty l}$ the
lower limit of the redshift interval at which the source is located, and $w_{\ssty k}(z)$ the
incompleteness correction for effective area of the source, corresponding to its computed flux,
as a function of the redshift.

Although these corrections are computed from local
simulations (z=0), and, therefore, they do not assume any cosmological model, the computed flux of each
source as a function of the redshift depends on
its luminosity distance, and therefore may change with the cosmological model assumed. In
addition to this implicit effect, the radial comoving distance $r$ and its redshift derivative,
$dr/dz$, also depend explicitly on the cosmology.

For each luminosity bin centred around $\bar{L}$ in each redshift interval centred around
$\bar{z}$, we compute the $1/V_{\ssty max}$ estimator for the luminosity function in that bin,
$\phi_{\ssty \bar{z},\bar{L}}$, as,
\begin{equation}
\lb{vmaxlf}
\phi_{\ssty \bar{z},\bar{L}} = \frac{1}{(\Delta L)_{\ssty \bar{L}}}
\sum_{i=1}^{N_{\ssty \bar{z},\bar{L}}} \frac{1}{V_{\ssty max}^{\ssty i}},
\end{equation}
where $(\Delta L)_{\ssty \bar{L}}$ is the length of the luminosity bin centred on $\bar{L}$,
and $N_{\ssty \bar{z},\bar{L}}$ the number of sources inside that luminosity bin and
redshift interval.

Assuming Poisson uncertainties, the error bars $\delta \phi$, can be estimated simply by,
\begin{equation}
\lb{vmaxlferr}
\delta \phi_{\ssty \bar{z},\bar{L}} = \frac{1}{(\Delta L)_{\ssty \bar{L}}}
\sqrt{ \sum_{i=1}^{N_{\ssty \bar{z},\bar{L}}}\left( \frac{1}
{V_{\ssty max}^{\ssty i}}\right)^2 }.
\end{equation}
Next, we briefly recall the properties of the void models used in the
computation of the LF, with a few key results needed in the interpretation of
the results.

\section{LTB/GBH dust models}
\citep[henceforth GBH]{2008JCAP...04..003G} have shown that an LTB
dust model could be parametrized to fit successfully and simultaneously many independent observations,
{ without the inclusion of a cosmological constant. The extra dimming of distant SNe Ia, as compared to
their expected observed fluxes in a flat, spatially homogeneous, Einstein-de Sitter (EdS) Universe, is then
understood not as being caused by an acceleration of the expansion rate, but rather as an extra
blueshift of the incoming light, caused by a non-homogeneous matter distribution in the line-of-sight}.
This so-called void model is characterised by an { effective} under-dense region of Gpc scale
around the Galaxy, { as opposed to the average spatial homogeneity supposed to hold at that scale by
the standard model. In this under-dense region}, both the matter density profile $\Omega_{\ssty M}$ and the
transverse Hubble constant $H_{\ssty 0}$ are functions of the radial coordinate $r$. 
At high enough redshifts though, the model is made to converge
to an EdS-like solution, making the non-homogeneity a localized property of the
model, and naturally reconciling it with the observed degree of isotropy in the cosmic microwave background
radiation maps. { The use of a pressure-less (dust) energy-momentum tensor, as opposed to the perfect
fluid one allowed in the standard model, is required in order to obtain an exact solution for Einstein's
field equations assuming the LTB line element. At early ages (high redshifts), radiation dominated
the Universe's energy budget, and the pressure term was relevant, but as discussed before, at these scales,
the LTB model is made to converge to the EdS solution by the GBH parameterization. At later ages
(low redshifts), radiation pressure is negligible, and so the use of a dust energy-momentum tensor is
well justified}. Geometrically, { the LTB dust model} is an analytical solution
for the Einstein's field equations, and arguably the simplest way to release the spatial homogeneity
assumption present in the standard model of cosmology.

{ Spatial homogeneity is, nevertheless, a symmetry assumption that greatly simplifies the model. Removing
it will unavoidably increase the degrees of freedom of the model. Because of that, even the most
constrained parameterizations of the LTB models still show an increased number of free parameters
when compared to the standard model. Since the quality of the combined fits to the observations these
alternative models can produce is similar to the one produced by the standard model, any analysis
that penalizes a greater degree of freedom of a model, like those used in \citep{2012JCAP...10..009Z} and
\citep{2012arXiv1208.4534D}, will disfavour such parameterization of the LTB model in comparison to
$\Lambda$CDM.

}

\subsection{Distances \& comoving volume}
\label{LTB}

Two quantities involved in the computation of the LF are affected
by a change in the cosmology: the luminosity of the sources, as computed in
equation (\ref{rflum}), through a change in the $\dl(z)$ relation, and their
enclosing comoving volume, as computed in equation (\ref{Vc}), through a change
in both $r(z)$ and $dr/dz$ relations. The aim is, therefore, to obtain those
last three relations in the constrained GBH (hereafter CGBH, see below)
{ model, that is the giant-void-GBH parameterization of the LTB dust model,
with best fit parameters from} \citep{2012JCAP...10..009Z}.

For comparison, we list the corresponding equations in the
$\Lambda$CDM standard model. Throughout this session, { where such comparisons
are made, we use the index $\Lambda$ on the left-hand side to identify an equation
computed in the standard model, and V to identify those obtained in the
CGBH model}.

We start by writting the Lema\^\i tre-Tolman-Bondi line element $\void{ds}$ in
geometrized units ($c$ = $G$ = 1) as,
\begin{equation}
\label{LTBmetric}
\void{ds}^2=-dt^2+\frac{A'(r,t)^2}{1-k(r)}dr^2+A(r,t)d\Omega^2,
\end{equation}
where $d\Omega$ is the spherical solid angle element, $A(r,t)$ the angular
diameter distance and $k(r)$ an arbitrary function that can be reduced to the
 Friedmann-Lema\^\i tre-Robertson-Walker line element $\lcdm{ds}$,
\begin{equation}
\label{FLRWmetric}
\lcdm{ds}^2 = -dt^2 + \frac{a(t)^2}{1-k\,r^2}dr^2+a(t)^2 r^2 d\Omega^2,
\end{equation}
by the suitable choice of homogeneity conditions, $A(r,t) = a(t) \, r$, and
$k(r) =\kappa \, r^2$, where $\kappa$ is the spatial curvature parameter, and $a(t)$
is the scale factor, both in the FLRW metric.

Because of the higher degree of freedom in the LTB metric (\ref{LTBmetric}), some
extra constraining conditions must be imposed. One of these conditions is that the big-bang
hyper-surface be constant in time coordinate, or, that the big-bang event occurred
simultaneously for all observers. This eliminates one degree of freedom of the
model.
The class of cosmological models with an LTB metric, a pressure-less (dust) content
distributed according to an under-dense matter profile $\Omega_{\ssty M}(r)$ around
the Galaxy, and simultaneous big-bang time is known as CGBH, that is, a constrained
case of the GBH void model.

The free parameters in the CGBH model are the expansion rate at the center of the
void $H_{\ssty in}$,
and the ones that characterize the matter density profile $\Omega_{\ssty M}(r)$:
the underdensity value at the center of the void, $\Omega_{\ssty in}$, the size
of the underdense region $R$, and the width of the transition $\Delta R$ between
the underdense interior and the assymptotic Einstein-de Sitter density
$\Omega_{\ssty out}$ at very large scales.
\citep{2012JCAP...10..009Z} consider both the case of an assymptotically flat
($\Omega_{\ssty out}$ = 1) Universe and that of an open ($\Omega_{\ssty out} \leq$ 1,
hereafter OCGBH), which they show to allow for a better fit to the cosmic
microwave background radiation (CMB).

\begin{figure}[htb]
\centering
\begin{tabular}{c}
\includegraphics[trim=8mm 2mm 6mm 7mm, clip, width=8.5cm]{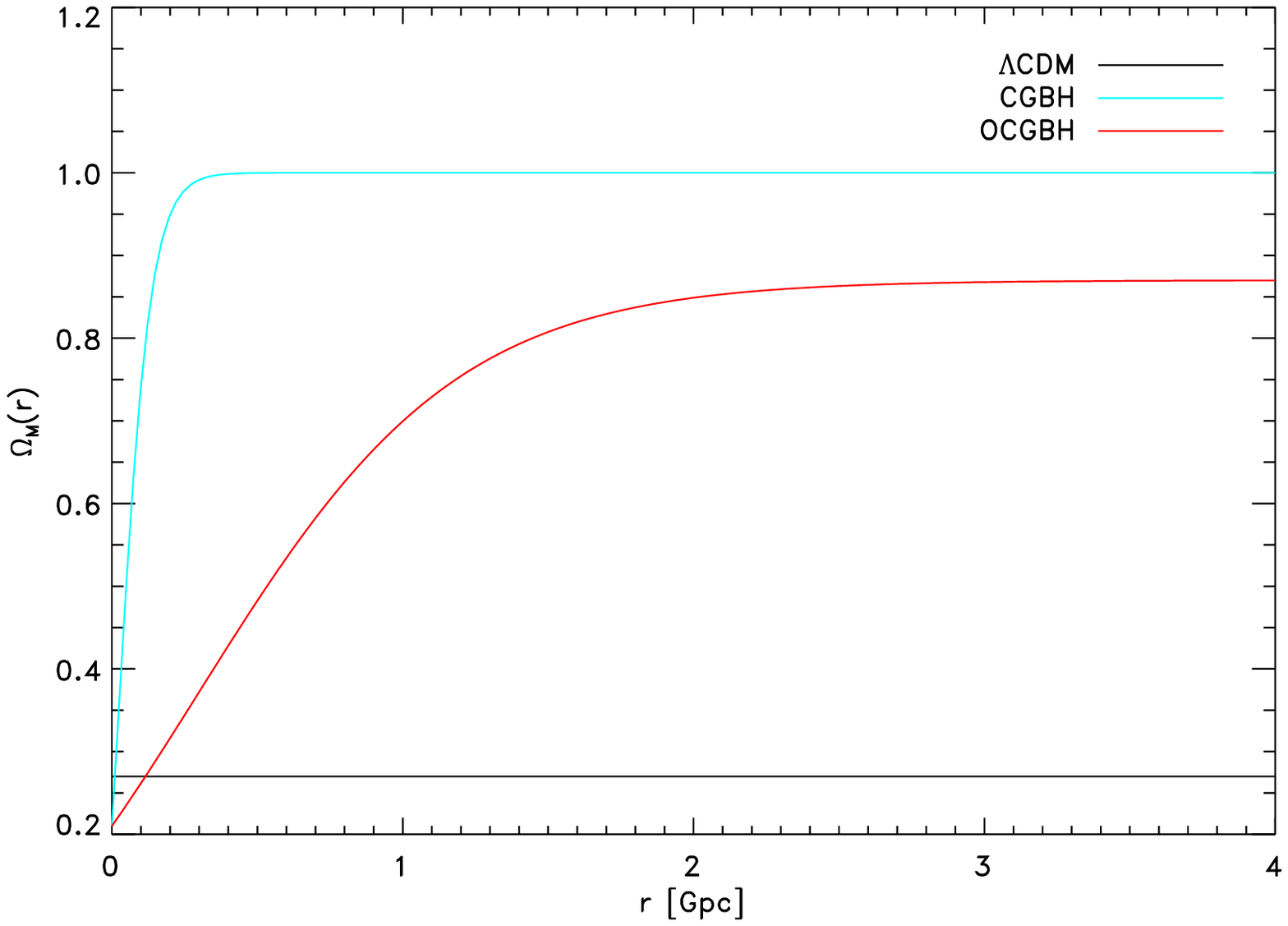} \\
\includegraphics[trim=8mm 2mm 6mm 7mm, clip, width=8.5cm]{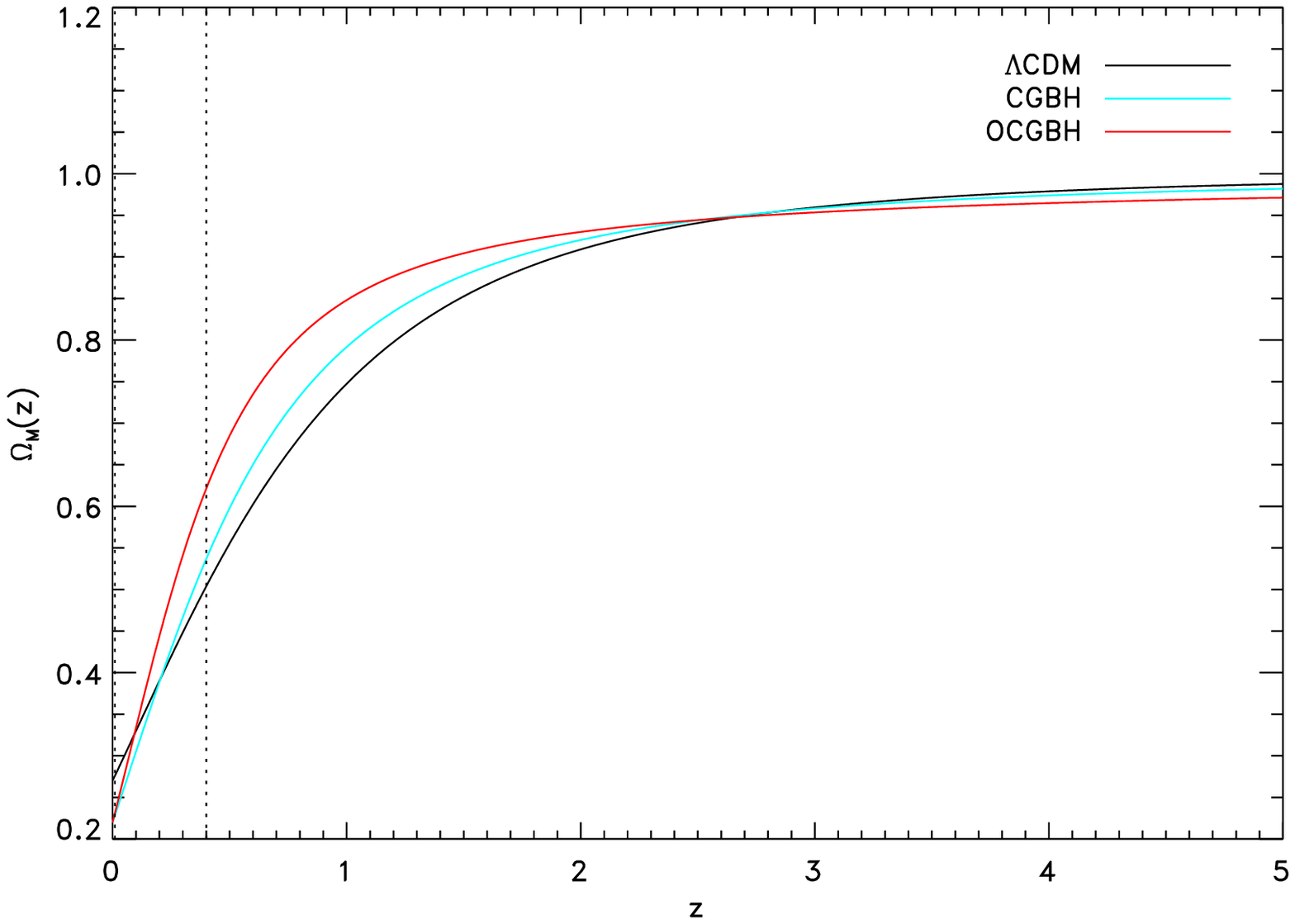}
\end{tabular}
\caption{{\em Upper panel:} Present time (t=t$_{\ssty 0}$) matter density parameters in the standard
($\Lambda$CDM, black line) and the void (GBH) cosmological models
(red and cyan lines). \label{plotom}
{\em Lower panel:} Redshift evolution of the dimensionless matter density parameters in the
standard ($\Lambda$CDM) and the void (GBH) cosmological models. The dotted vertical lines delimit
the lowest redshift interval considered in the computation of the LF, at which the faint-end
slopes are fit. \label{plotomz}}
\end{figure}

The matter density profile $\Omega_{\ssty M}(r)$ is written in the GBH model as a
function of the fit parameters as,
\begin{equation}
\label{GBHOM}
\Omega_{\ssty M}(r) = \Omega_{\ssty out} + (\Omega_{\ssty in}-\Omega_{\ssty out})
\left( \frac{1-\tanh[(r-R)/2\Delta R]}{1+\tanh[R/2\Delta R]} \right).
\end{equation}
whereas the present time transverse Hubble parameter $H_{\ssty 0}(r)$ is,
\begin{equation}
\label{GBHH0}
H_{\ssty 0}(r) = H_{\ssty in} \left[ \frac{1}{\Omega_{\ssty k}(r)} -
\frac{\Omega_{\ssty M}(r)}{\Omega_{\ssty k}(r)^{3/2}}\sinh^{-1}
\sqrt{\frac{\Omega_{\ssty k}(r)}{\Omega_{\ssty M}(r)}} \right],
\end{equation}
with $\Omega_{\ssty k}(r) = 1-\Omega_{\ssty M}(r)$, { the curvature parameter
inside the under-dense region, needed to close the Universe}. In the standard model
both matter density parameter and the Hubble parameter do not depend on the radial
coordinate. In all the standard model computations done in this work we use
$\Omega_{\ssty M} = 0.27$, $\Omega_{\Lambda} = 0.73$, and
$H_{\ssty 0} = 71$ km s$^{-1}$ Mpc$^{-1}$ { as obtained in \citep{2011ApJS..192...18K}}.
Figure \ref{plotom} shows
the comparison of the evolution of the matter density parameter in the
standard and in the void cosmologies.

With those definitions, the angular diameter distance $A(r,t)$ can be computed in
parametric form as,
\begin{equation}
\label{GBHA}
A(r,t) = \frac{\Omega_{\ssty M}(r)}{2[1-\Omega_{\ssty M}(r)]^{3/2}}[\cosh(\eta)-1] \,
A_{\ssty 0}(r),
\end{equation}
where $A_{\ssty 0}(r)$ is the angular diameter distance at $t=t_{\ssty 0}$, and
the parameter $\eta$ advances the solution given $r$, $t$, $H_{\ssty 0}(r)$, and
$\Omega_{\ssty M}(r)$ as follows,
\begin{equation}
\sinh(\eta) - \eta = 2 \, \frac{[1 - \Omega_{\ssty M}(r)]^{3/2}}{\Omega_{\ssty M}(r)}
\, H_{\ssty 0}(r) \, t.
\end{equation}
Once the angular diameter distance $\da(z) = A[r(z),t(z)]$, is computed, we can use the
very general reciprocity theorem \citep{1933PMag...15..761E},
\begin{equation}
\label{rec}
\dl = (1+z)^2 \da = (1+z) \dg,
\end{equation}
to compute the luminosity distance $\dl(z)$ in the void models considered here. In
the equation above, $\dg$ is the {\em galaxy area distance}, which reduces to the comoving
distance in FLRW models.
We still need to obtain the $r(z)$ and $t(z)$ relations in this Cosmology. We start with
the radial null-geodesic equation, which can be written by making $ds^2=d \Omega ^2=0$,
yielding,
\begin{equation}
\bigvoid{\frac{dt}{dr}} = -\frac{A'(r,t)}{\sqrt{1-k(r)}},
\end{equation}
where the minus sign is set for incoming light. The corresponding standard model
equation reads,
\begin{equation}
\biglcdm{\frac{dt}{dr}} = -\frac{a(t)}{\sqrt{1-\kappa r^2}},
\end{equation}
In the LTB metric, the time coordinate-redshift relation can be obtained to first order
in wavelength starting from the redshift definition,
as in,  \citep[e.g.][]{2007JCAP...02..019E},
\begin{equation}
\bigvoid{\frac{dt}{dz}} = - \frac{1}{1+z}\frac{A'}{\dot{A}'},
\end{equation}
whereas the corresponding $\Lambda$CDM relation can be written as,
\begin{equation}
\biglcdm{\frac{dt}{dz}} = - \frac{1}{1+z}\frac{a}{\dot{a}}.
\end{equation}

The last two equations allow us to write the radial coordinate
$r$ in terms of the redshift $z$, by solving the following expression,
\begin{equation}
\label{drdz}
\bigvoid{\frac{dr}{dz}}=\frac{1}{1+z}\frac{\sqrt{1-k(r)}}{\dot{A}'},
\end{equation}
where $k(r)$ can be written in terms of the formerly defined quantities as,
\begin{equation}
\label{kr}
k(r) = -\Omega_{\ssty k} \, H_{\ssty 0}^2(r) \, r^2.
\end{equation}
Similarly, we can write, again for comparison,
\begin{equation}
\label{LCDMdrdz}
\biglcdm{\frac{dr}{dz}}=\frac{1}{1+z}\frac{\sqrt{1-\kappa r^2}}{\dot{a}}.
\end{equation}

It is worth noting that the comoving distance $\dc$ is not, {\em in general}, equal
to the galaxy area distance, $\dg$, related to the luminosity distance through the
reciprocity theorem (\ref{rec}).

The usual comoving-to-luminosity distance relation, \textbf{$\dg = (1+z) \, \da = \dc$},
is only valid in the FLRW metric, for which the following relations holds,
\begin{eqnarray}
\lcdm{(1+z)} = \frac{a_{\ssty 0}}{a(t)}, \\
d_{\ssty A,\Lambda} = r \, a(t),
\end{eqnarray}
where $a(t)$ is the usual scale factor, and $a_{\ssty 0}$, its value at present time
(t=0). The last equation is valid only if we set $a_{\ssty 0}=1$.
As a consequence, an LTB model with its luminosity distance-redshift relation
constrained to fit
the Hubble diagram for SNe Ia could still yield comoving distances, and therefore
volumes, significantly different then those obtained in the standard model.


The additional constraint imposed by the measurements of the characteristic angular size
of the Baryonic Accoustic Oscillations (BAO){, \citep[e.g.][]{2010MNRAS.401.2148P, 2012MNRAS.426.2719R},}
appears to pin down the comoving distance
quite effectively up to intermediate redshifts, and it turns out that the difference in
such distances computed in the $\Lambda$CDM and the GBH models is never larger than
10\% at $z = 1$. However, $\dc$ computed in the CGBH model at $z = 5$ is approximately
12\% smaller, and $\approx$ 17\% in the open CGBH one.

The non-linear nature of the equations relating distances to volumes and
luminosities, in particular for high redshift sources, must also be considered.
At redshift $z=0.4$,
for example, the luminosity distances computed in the void models, CGBH and OCGBH respectively,
are 4.90\% and 0.92\%  shorter then the standard model ones, whereas the comoving distances
in the void models are 4.88\% and 0.96\% shorter when compared to the standard model value.
Such differences correspond to an extra dimming in luminosities equal to 10.04\% in the CGBH
model and 1.85\% in the OCGBH one. The corresponding shrink in volumes are 15.35\% and 2.84\%,
for the CGBH and OCGBH models.

\begin{figure}[htb]
\centering
\begin{tabular}{c}
\includegraphics[trim=0mm 14mm 4mm 6mm, clip, width=8.5cm]{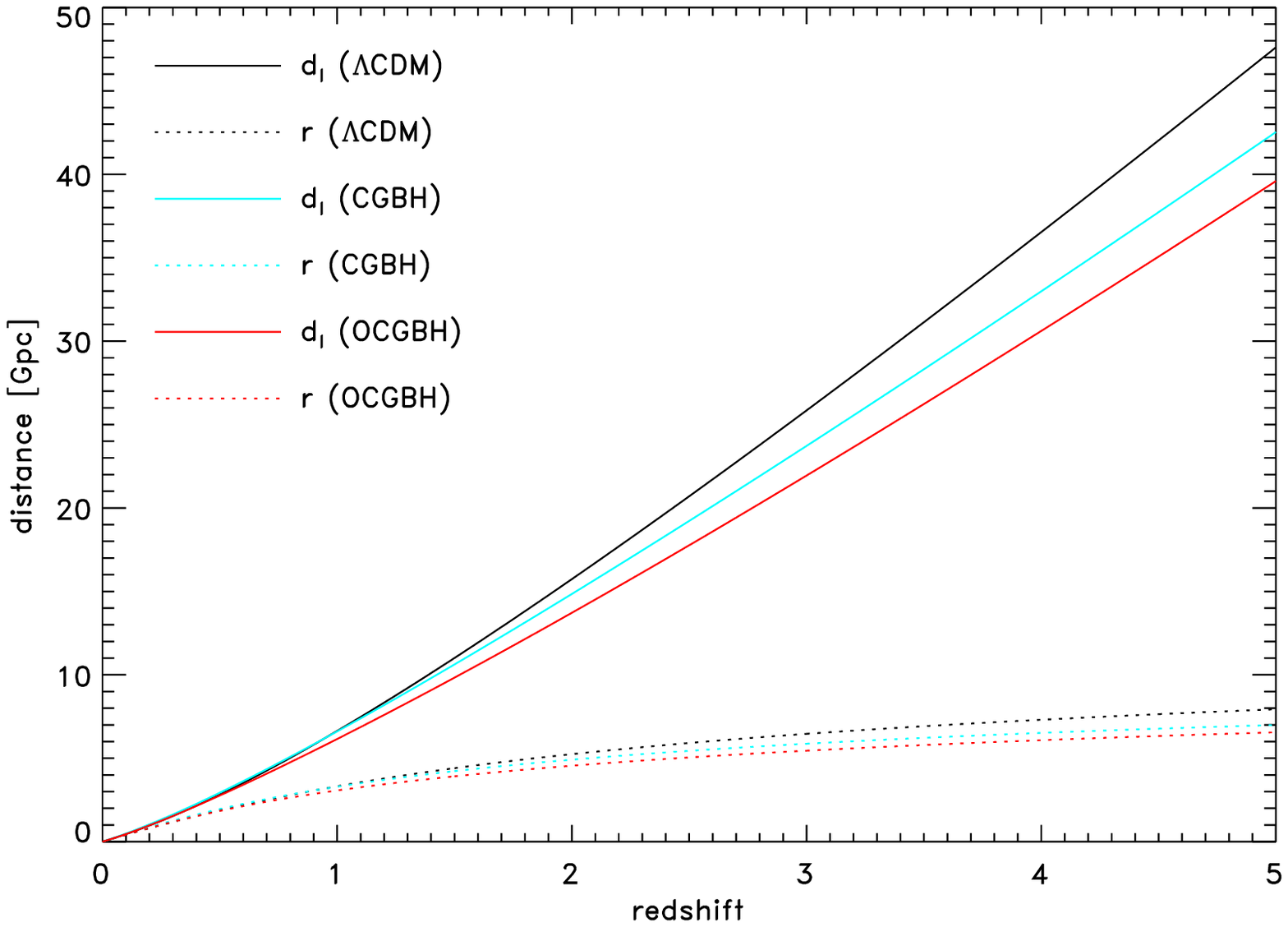} \\
\includegraphics[trim=0mm 4mm 4mm 6mm, clip, width=8.5cm]{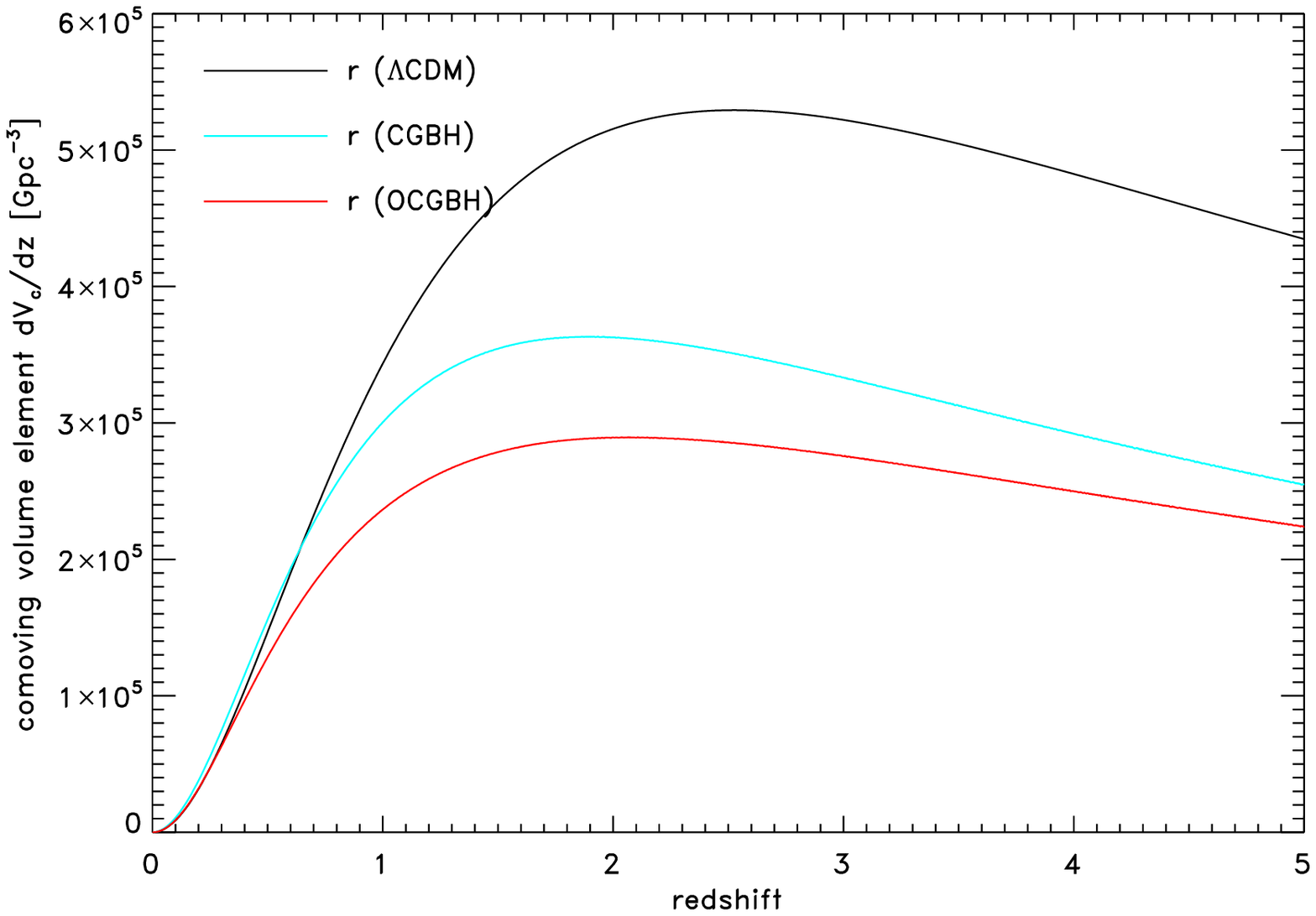}
\end{tabular}
\caption{{\em Upper panel:} Luminosity (solid lines) and comoving distances (dotted lines) versus the redshift
in the standard ($\Lambda$CDM) and the void (GBH) cosmological models. Up to redshift $z \approx 1$, both distances
in the constrained, flat void model (CGBH) follow very closely their Standard Model counterparts, but even in
the case of the best fit parameters in \citep{2012JCAP...10..009Z} yield increasingly different distances with
the redshift.
\label{plotdist}
{\em Lower panel:} Comoving volume elements in the standard ($\Lambda$CDM) and the void (GBH)
cosmological models. The quantities in the void models adopted here, evolve with the
redshift in a similar way as in the standard model, but at redshifts higher then
approximately 0.6 their values are consistently lower then in the $\Lambda$CDM model.
\label{plotvol}}
\end{figure}

Such non-linearities that can make small discrepancies in luminosity and comoving distances
caused by the central underdensity in GBH models sum up to non-negligible differences
{\em in the shape of the LF}, when compared to the standard model.
This can be understood by looking at Figure \ref{plotdist}, where the luminosity
distance and the comoving distance are plotted against the redshift. 

For any given redshift $z'$, consider the differences $\Delta \dl(z')$ and $\Delta r(z')$ 
between the distances computed in the standard model and those in the void models. Both
differences depend on the redshift and do not, in general, cancel out or even yield
a constant volume-to-luminosity ratio as a function of the redshift. As a result, the
number of sources in each luminosity bin might change due to differences in the
luminosities.

Additionally, the weight 1/V$_{\ssty max}$ that each source add to the LF in
that bin will not be the same, leading to a LF value in that luminosity bin in the void
model that is different then the one in the standard model, even if the sources inside
the bin are the same. Figure \ref{plotvol} shows the comoving volume element in the different
cosmologies adopted here.

Such differences in the estimated value for the LF in each luminosity bin will not,
in general, be the same. As a consequence, not only the normalization but also the shape of
the LF might change from one cosmology to another.


\section{Results}
\label{results}
We compute the rest frame monochromatic and total IR luminosity LF for sources in
the combined fields, blind selected in the 100 $\mu$m and 160 $\mu$m bands, using
the non-parametric $1/V_{max}$ method, both in the Standard Model and the GBH void
models. The LF values are listed in Tables
\ref{vmt100LCDM}-\ref{vmtLIR160oGBH}, up to redshifts $z \approx 3$
for the monochromatic LFs and up to $z \approx 4$ for the total IR
ones.




We use the same binning in luminosity and redshift as in \citep{2013MNRAS.tmp.1158G}.
The average values for the redshift intervals are 0.2, 0.6, 1.0, 1.5, 2.1,
and 3.0, for the monochromatic LFs, and 0.2, 0.4, 0.5, 0.7, 0.9,
1.1, 1.5, 1.9, 2.2, 2.8 and 3.6 for the total IR ones. The effective wavelengths
will be 60 and 90 $\mu$m in the rest-frame LF.
Due to the lack of enough 1/V$_{\ssty max}$ LF points to fit a Schechter function in
the higher redshift bins, our analyses of the monochromatic luminosity functions is
limited to intervals $\bar{z} \leq 3$. As a consistency check we compared our results
for the standard model with Gruppioni et al.'s and the agreement is excellent.

The monochromatic and the total luminosity LFs are shown in Figures
\ref{plot100}-\ref{plotLIR160} for the three cosmologies considered in this paper.
When comparing the values in each luminosity bin it is clear that in void cosmologies the density is lower at the
lowest luminosities. While at redshift larger than 0.8 the incompleteness at low luminosities does not allow
to draw any firm conclusion on this, in the two lowest redshift bins, the void models show LF values up to an order of magnitude
lower than their $\Lambda$CDM counterpart, at L$\leq$10$^{10}$L$_\odot$.


The resulting differences in the LF computed in different models 
show up at the faint luminosity end of the luminosity functions. We use the Schechter's analytical profile \citep{1976ApJ...203..297S},
\begin{equation}
\lb{schfc}
\varphi(L) = \frac{\phi^{\ssty *}}{L^{\ssty *}} \,
\left(\frac{L}{L^{\ssty *}}\right)^{\alpha} \, e^{-L/L^{\ssty *}} = 
\varphi^{\ssty *} \, \left(\frac{L}{L^{\ssty *}}\right)^{\alpha} \,
e^{-L/L^{\ssty *}},
\end{equation}
and fit it to the $1/V_{max}$ points over the ($\bar{z}$,$\bar{L}$) bins, where $\phi^{\ssty *}$ is
the comoving number density normalisation, $L^{\ssty *}$ is the characteristic luminosity
and $\alpha$ is the faint-end slope, using the IDL routine MPFITFUN
\citep{2009ASPC..411..251M}, based on the Levenberg-Marquardt algorithm
\citep{1978LNM...630...105T}. { For each best fit parameter, its formal 1-$\sigma$ uncertainty
is obtained by taking the square root of its corresponding element in the diagonal of the
3x3 covariance matrix of the fitting procedure \citep[see also][]{1995TDAPR.122..107R}.}

Since we are primarily interested in checking possible changes in the LF caused by the underlying
cosmologies, we chose to use the classical Schechter function, instead of the double exponential
one \citep{1990MNRAS.242..318S}. The latter better fits the FIR LF bright-end,
but the Schechter function have { fewer} free parameters, which allows it to fit higher redshift
intervals, where the number of data points is small.

First we check for any variation of the $\alpha$ parameter with redshift, and find that it is
consistent with no evolution. 
We test the incompleteness using the $V_e/V_a$ tests \citep{1980ApJ...235..694A}. A given
($\bar{z}$,$\bar{L}$) bin is considered complete by this test if its $V_e/V_a$ value is
1/2. We find that the 1/V$_{\ssty max}$ LF points do not suffer from significant incompleteness
at $\bar{z}=0.2$ where the $V_e/V_a$ values in lowest luminosity bins of the monochromatic
luminosity functions are 0.6 $\pm$  0.1 and 0.5 $\pm$ 0.1 for the rest-frame 100 and 160
$\mu$m, respectively. Such values become 0.15 $\pm$ 0.09 and 0.22 $\pm$ 0.03 at $\bar{z} = 1$,
and 0.12 $\pm$ 0.05 and 0.11 $\pm$ 0.04 at $\bar{z}=3$.


This is because at higher redshifts the flux limit of the observations corresponds to
increasingly different luminosity limits, depending on the SED of the sources, leading
to an incompleteness in the lower luminosity bins that is dependent on the galaxy type
\citep{2004MNRAS.351..541I}.
Because of this, in the fits presented in Tables \ref{scht100}-\ref{schtLIR160}, we
chose to fix the $\alpha$ parameter to its value in the lower redshift interval.

\section{Discussion}
\label{discussion}
\begin{figure*}
\centering
\includegraphics[width=17cm]{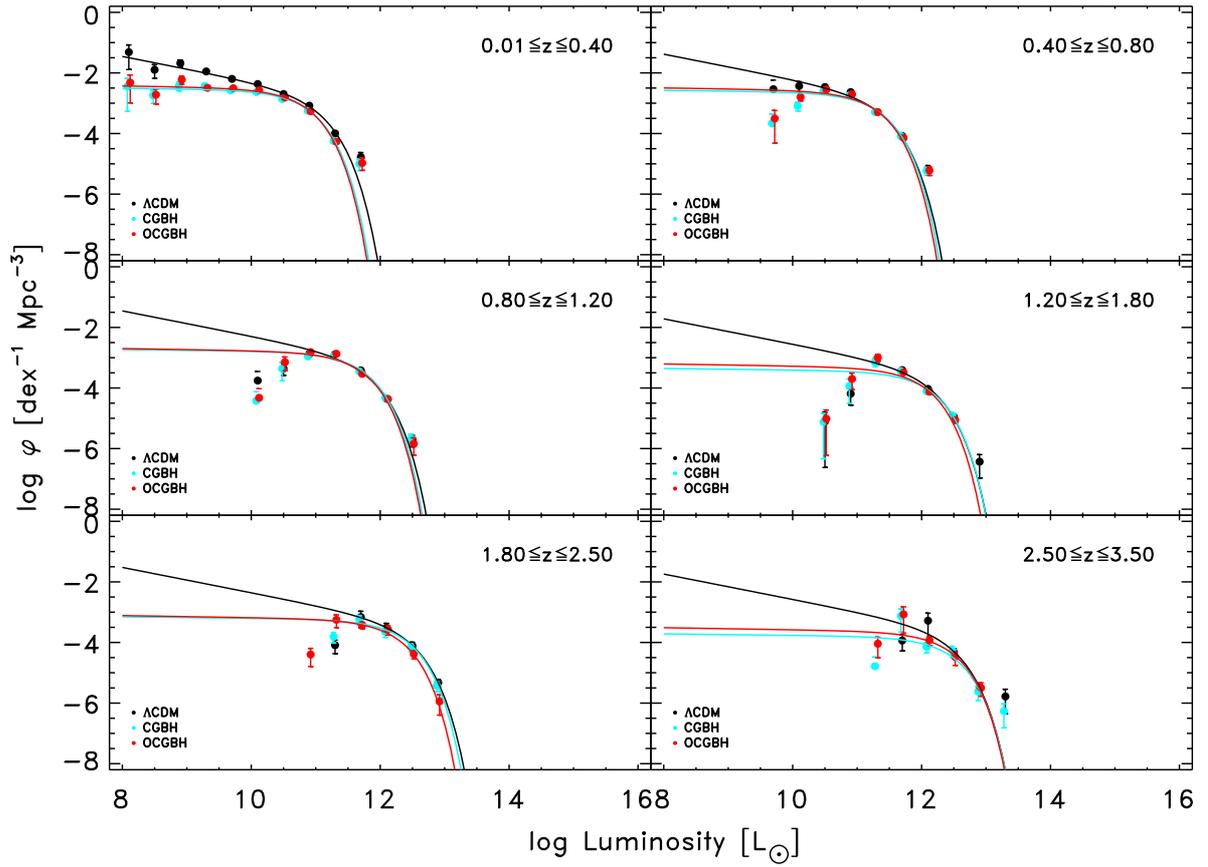}
\caption{Luminosity functions derived in the standard ($\Lambda$CDM) (black dots) and the void (GBH)
cosmological models (red and cyan dots). We show also the best-fit Schechter profiles to the rest-frame
100 $\mu$m 1/V$_{\ssty max}$ corresponding to effective wavelengths of 60 $\mu$m. \label{plot100}}
\end{figure*}

\begin{figure*}
\centering
\includegraphics[width=17cm]{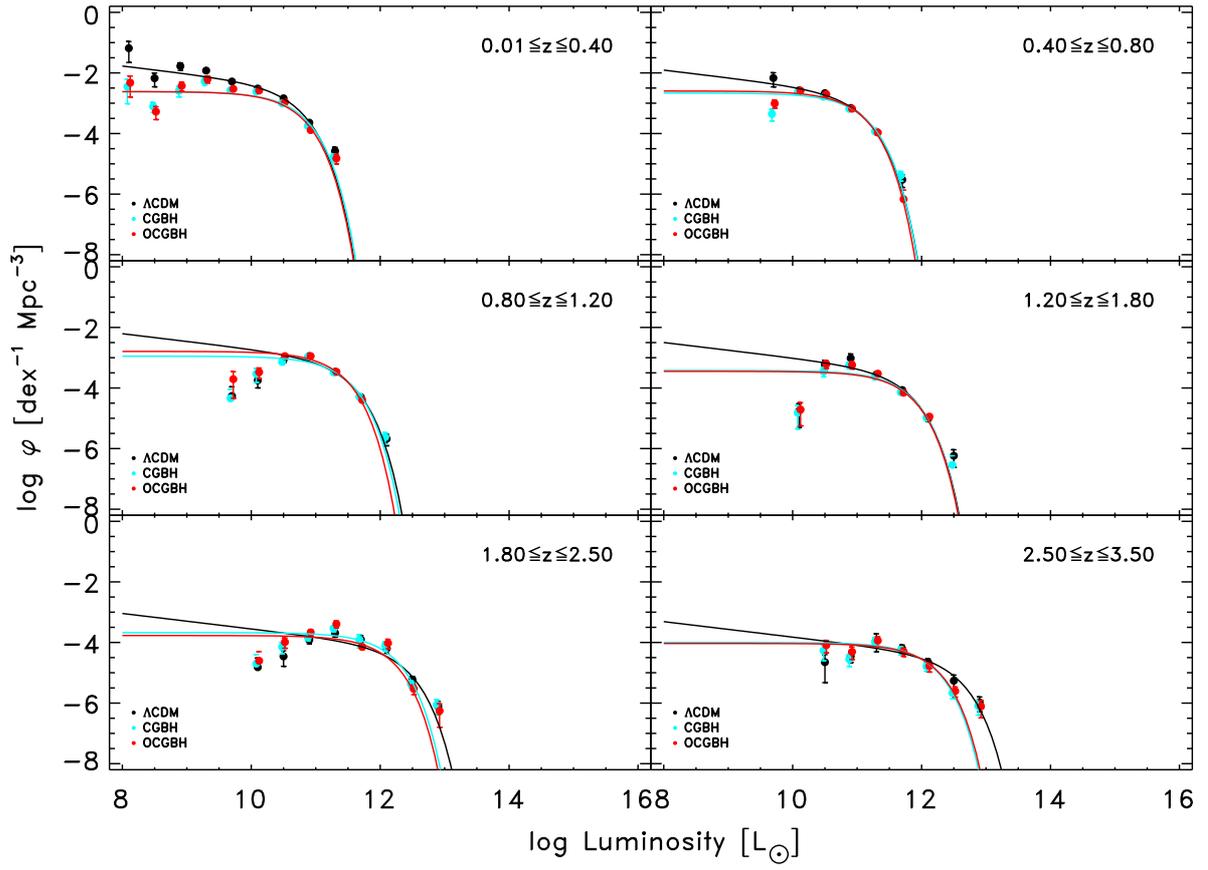}
\caption{As Figure \ref{plot100} for the rest-frame 160 $\mu$m 1/V$_{\ssty max}$
luminosity functions. Here the effective wavelength is 90$\mu$m. \label{plot160}}
\end{figure*}

\begin{figure*}
\centering
\includegraphics[width=17cm]{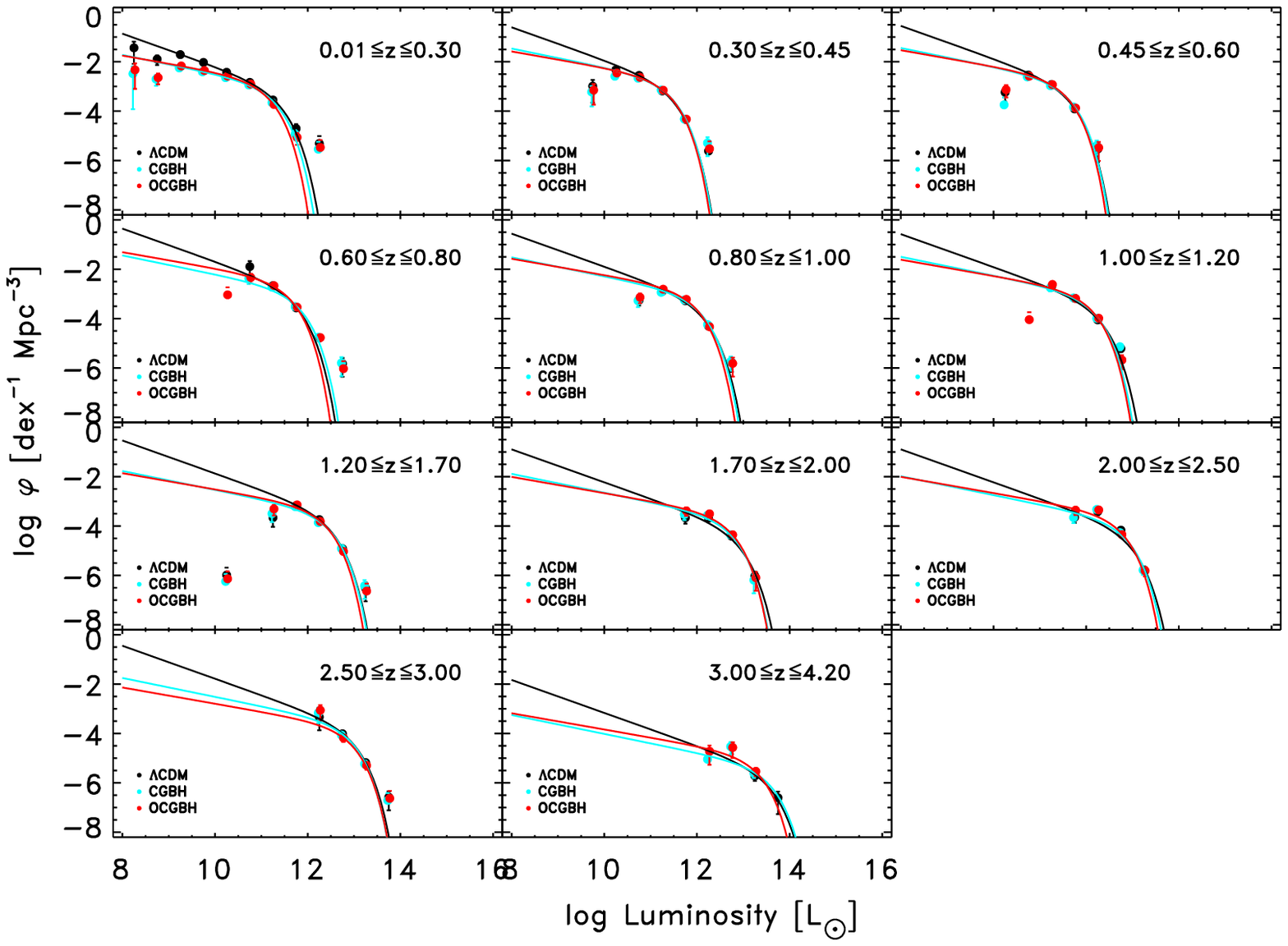}
\caption{Schechter profile fits to the rest-frame total IR luminosity functions computed from
the PACS 100 $\mu$m 1/V$_{\ssty max}$ band, assuming the standard ($\Lambda$CDM)
and the void (GBH) cosmological models. \label{plotLIR100}}
\end{figure*}

\begin{figure*}
\centering
\includegraphics[width=17cm]{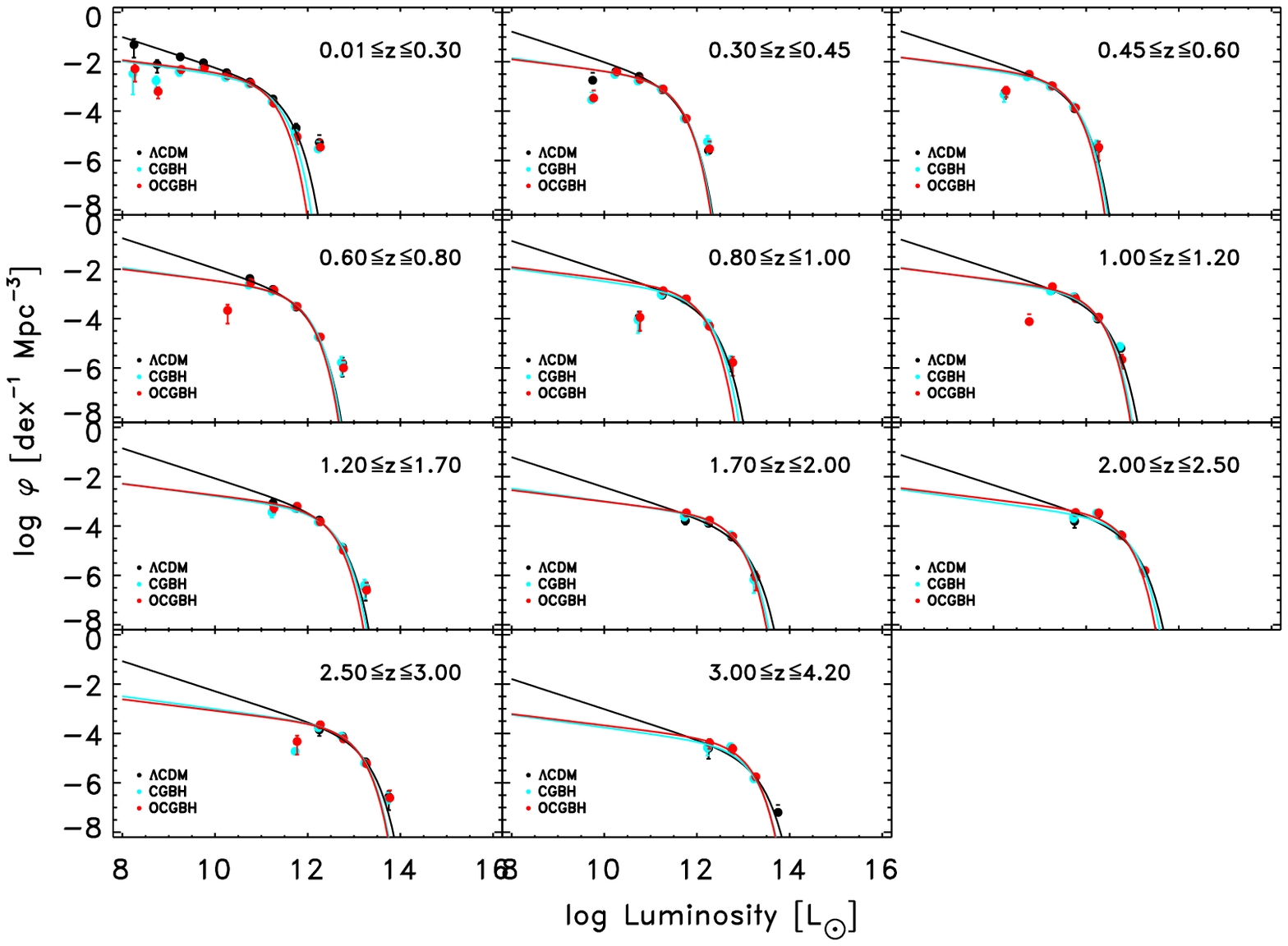}
\caption{Schechter profile fits to the rest-frame total IR luminosity functions computed from
the PACS 160 $\mu$m 1/V$_{\ssty max}$ band, assuming the standard ($\Lambda$CDM)
and the void (GBH) cosmological models. \label{plotLIR160}}
\end{figure*}

In Figures \ref{plot100}-\ref{plotLIR160} we plot the $1/V_{\ssty max}$ LF
estimations in the three different cosmologies, together with the best fit Schechter profiles
for each of them. As can be seen on the four Figures, the faint end number densities in
the void models are lower then the standard model ones.

If there was a direct correlation
between the matter density parameter in the cosmology, and its estimated number density of
sources selected in the FIR, then at the lowest redshift bin we should see higher number
densities in the void models, since the $\Omega_{\ssty M}(z)$ in those models are bigger in
that redshift range then the standard model value (more on that in Appendix \ref{mock},
Figure \ref{plotomz}).
The difference in the number densities at the lower redshift interval, for the different
cosmologies, not only does not follow the same relation as the matter density parameters
$\Omega_{\ssty M}(z)$, but also shows a dependence on the luminosity, being more pronounced
at the fainter end in both the monocromatic and the total IR luminosity LFs.
Such dependence produces significant differences in the faint-end slopes of
the computed luminosity functions. That can only be attributed to the different
geometrical parts of the cosmological models studied here, since the matter
content, as discussed above, would only shift the normalization of the LF,
independently of the luminosity of the sources.

In Table \ref{alphat} we present the best fit values of $\alpha$
for each dataset / model combination.
{ Simple error propagation allow us to write the uncertainty of the difference
$\Delta \alpha$, between the faint-end slopes in the standard, $\alpha_{\ssty \Lambda}$,
and the void, $\alpha_{\ssty V}$, models as,
\begin{equation}
\delta (\Delta \alpha) = \sqrt{(\delta \alpha_{\ssty \Lambda})^2 \, + \, (\delta \alpha_{\ssty V})^2}.
\end{equation}
The significance level of such difference can then be obtained by computing
$\Delta \alpha / \delta(\Delta \alpha)$. For the monochromatic 100-$\mu$m luminosity
functions, the difference between the $\alpha$ computed assuming the standard model
and the ones computed in the GBH models studied here is 6.1-$\sigma$, as compared to its propagated
uncertainty. For the monochromatic 160-$\mu$m luminosity functions, this value is
3.2-$\sigma$. For the total-IR 100-$\mu$ selected luminosity functions, these values
are 2.9-$\sigma$ for the difference between $\Lambda$CDM and CGBH models, and 3.1-$\sigma$,
for the $\Lambda$CDM-OCGBH difference. Finally, for the same differences
in the total-IR 160-$\mu$m selected dataset, the significances are 3.1-$\sigma$, and
3.4-$\sigma$, respectively.}


Could this difference be caused by a limitation of the $1/V_{\ssty max}$ method used here?
As we show in the Appendix \ref{mock}, the matter density parameters in both
$\Lambda$CDM and GBH models do not affect the performance of this LF estimator significantly:
when the same input LF and matter density are assumed, the 1/V$_{\ssty max}$ method obtains
values within their error bars for all cosmological models considered. This indicates
that this change in the slopes is caused by how the luminosities are computed from the
redshifts in the different metrics.

\begin{figure}[htb]
\centering
\begin{tabular}{c}
\includegraphics[trim=8mm 14mm 6mm 7mm, clip, width=8.5cm]{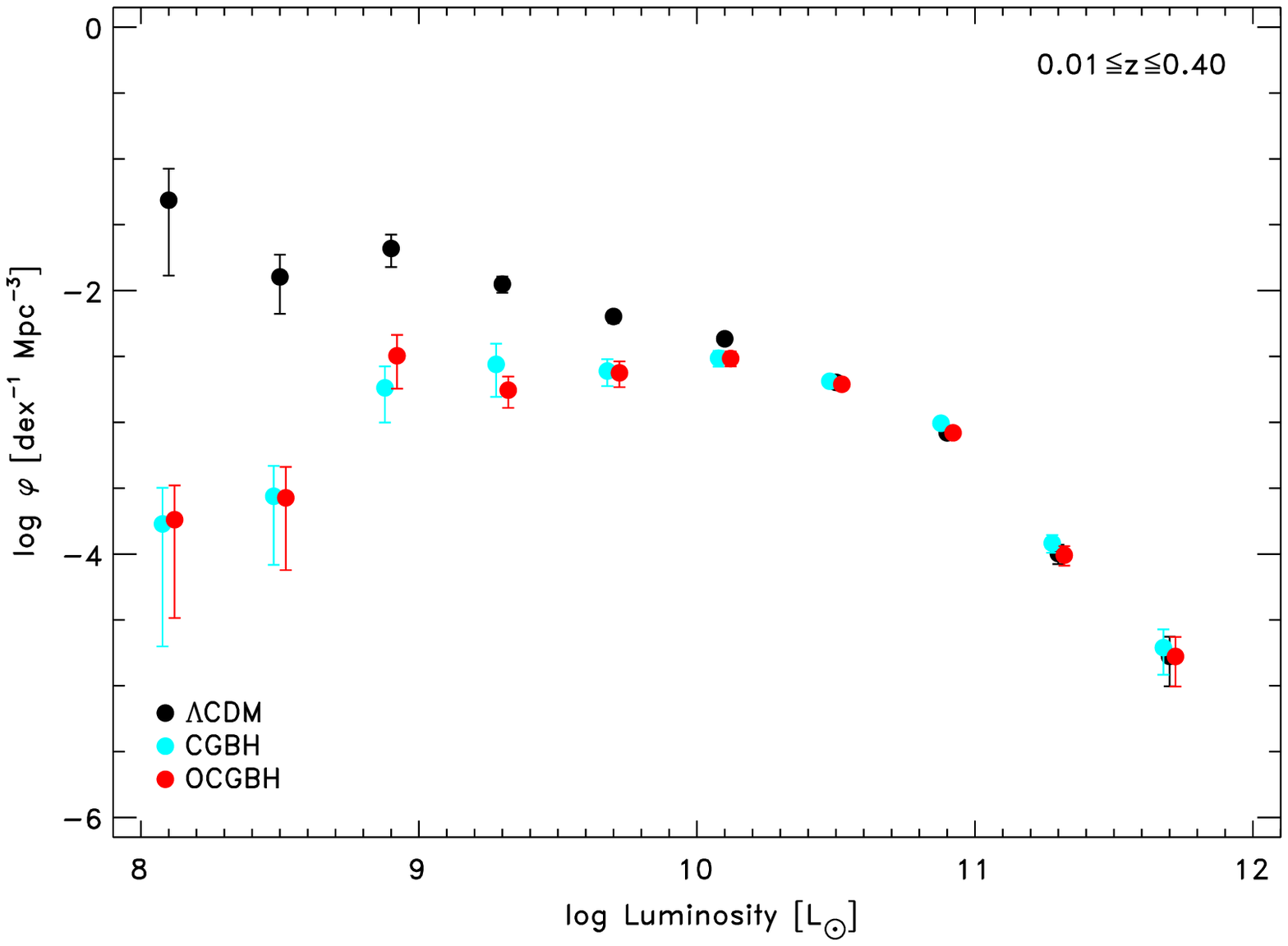} \\
\includegraphics[trim=8mm 0mm 6mm 7mm, clip, width=8.5cm]{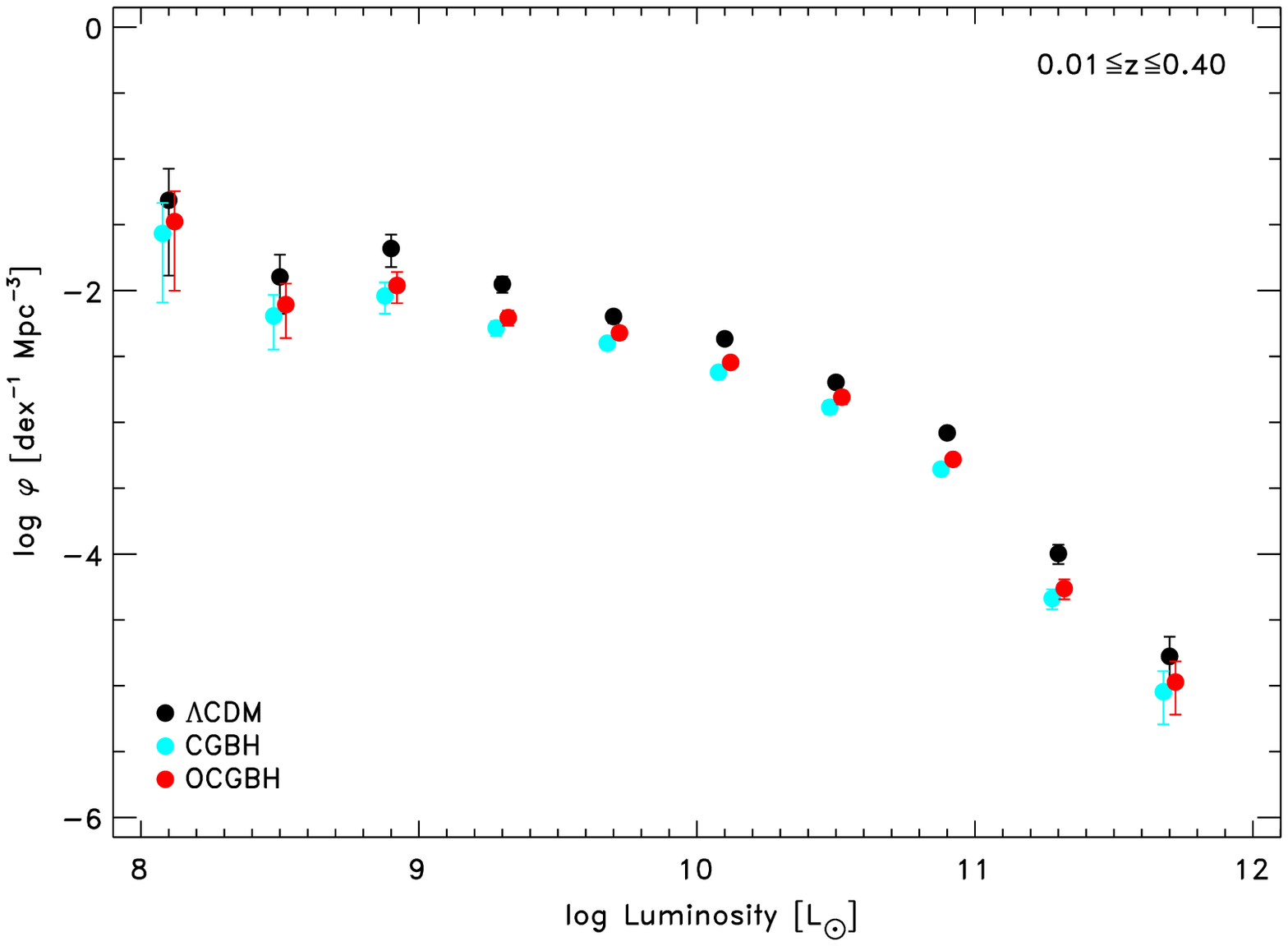}
\end{tabular}
\caption{{\em Upper panel:} Effect of the luminosity distance - redshift relation on the shape of the
LF. The black points were computed using the $1/V_{\ssty max}$ method,
assuming both $\dl(z)$ and $\dc(z)$ relations stemming from the standard model, while
the blue and red points kept the $\dc(z)$ relation for the $\Lambda$CDM model, changing
only the $\dl(z)$ relation for that in the listed void model. The effect of the $\dl(z)$
relation on the shape of the LF is clear, especially at the lower luminosity bins.
\label{plotLtest}
{\em Lower panel: }Effect of the comoving distance - redshift relation on the shape of the
LF. The black points were computed using the $1/V_{\ssty max}$ method,
assuming both $\dl(z)$ and $\dc(z)$ relations stemming from the standard model, while
the blue and red points kept the $\dl(z)$ relation for the $\Lambda$CDM model, changing
only the $\dc(z)$ relation for that in the listed void model. The effect of the $\dc(z)$
relation on the shape of the LF is found to be much less relevant then that of the
$\dl(z)$ one.
\label{plotVtest}}
\end{figure}

To further check this assertion, we investigate the effects of both luminosity and
comoving volume separately on the shape of the LF. Starting from the
$1/V_{\ssty max}$ results for the LF in the interval 0 $< z < $ 0.4, assuming the
standard cosmological model, we compute alternative LFs using the same methodology,
but assuming either the luminosity distance of one of the void models and keeping
the comoving distance of the standard model, or the luminosity distance of the
standard model and the comoving distance of one of the void models. This allows us
to assess how each distance definition affect the LF individually. The results are
plotted in Figure \ref{plotLtest}, for the rest frame
100-$\mu m$ monochromatic luminosity dataset.

From those plots it is clear to see that the luminosity is the main cause of
change in the shape of the LF. What remains to be investigated is whether the
number density in a given luminosity bin is lower in the void models because
of a re-arranging of the number counts in the luminosity bins, or because of
a possible change in the maximum volume estimate of the sources in each bin.
From this inspection,
it turns out that the number counts in all three models are all within their
Poisson errors, and therefore, the number densities in the void models are
lower because the maximum volumes in them are larger.

Looking at equation (\ref{Vc}), we identify two parameters that can
introduce a dependency of the maximum volume of a source on its luminosity:
the incompleteness corrections, $w_{\ssty k}(z)$, and the upper limit of the
integral, $z_{\ssty max}$.

The incompleteness correction, $w_{\ssty k}(z)$, for each source, depends on
the observed flux that source would have at that redshift, which is affected
by the luminosity distance-redshift relation assumed.

More importantly though,
at the higher luminosity bins, the $z_{\ssty max}$ of most of the sources
there assumes the $z_h$ value for that redshift interval, which does not depend
on the luminosity of the source. This renders the $V_{\ssty max}$ of the high
luminosity sources approximately the same, apart from small changes caused by
the incompleteness corrections $w_{\ssty k}$, as discussed above. At the lower
luminosity bins, on the other hand, it happens more often that the $z_{\ssty max}$
of a source assumes its $\zeta$ value, which in this case depends on its
luminosity, as it is clear from equation (\ref{zmaxdef}).

In order
for that equation to hold, given that $f_{\ssty R}$ and $f_{\ssty R,lim}$ are fixed,
the ratio between $\dl(z) / \dl(\zeta)$ must be the same for all cosmologies.
Since the redshift $z$ of each source is also fixed, then it follows that the
$\zeta$ value that makes the $\dl(z) / \dl(\zeta)$ ratio hold in the void models
must be higher then in the standard model (see Figure \ref{plotdist}). This, in
turn, accounts for the larger maximum volumes, and lower number densities, at
the low luminosity bins in the void models.

From the discussion above we conclude that a change in the luminosity distance - redshift
relation changes the $z_{\ssty max}$ of the low luminosity sources, which in turn
changes the maximum volumes, and finally, the fitted faint-end slope. However, from
Figure \ref{plotdist}, it is not obvious that such small differences in the $\dl(z)$
relation for the different cosmological models could cause such a significant
change in the faint-end slopes, especially at low redshifts.
It is useful to remind here that the LF is a non-linear combination of quantities
that depend, from a geometrical point of view, on the luminosity distance (through the
luminosities of the sources) and on the comoving distances (through their enclosing
volumes).
Even if the observational constraints on the luminosity-redshift relation, and
the additional ones stemming from BAO results, yield both $\dl(z)$ and $\dc(z)$ quite
robust under changes of the underlying cosmological models, such small differences
in the distances could pile up non-linearly,
and cause the observed discrepancies in the faint-end slope.

This appears to be the case here, at least in the low redshift interval where we
can fit the faint-end slopes with confidence. Rather then following the trend of
the matter density parameter, the number densities at those redshifts seem to
be predominantly determined by their enclosing volumes (even if at low redshift
the differences in the distance-redshift relations in the different cosmologies
is quite small).

Looking at how the distances in Figure \ref{plotdist} have increasingly different
values at higher redshifts, it would be interesting to check if the faint end slopes
in the different cosmologies at some point start following that trend.
Unfortunately, at higher redshifts the incompleteness caused by different luminosity limits
for different populations does not allow us to draw any meaningful conclusion about the
faint-end slope of the derived LFs. 
As it is, all that can be concluded is that the standard model LF would be over-estimating
the local density of lower luminosity galaxies if the Universe's expansion rate and history
followed that of the LTB/GBH models.

\begin{figure}
\centering
\begin{tabular}{c}
\includegraphics[trim=0mm 0mm 0mm 0mm, clip, width=9cm]{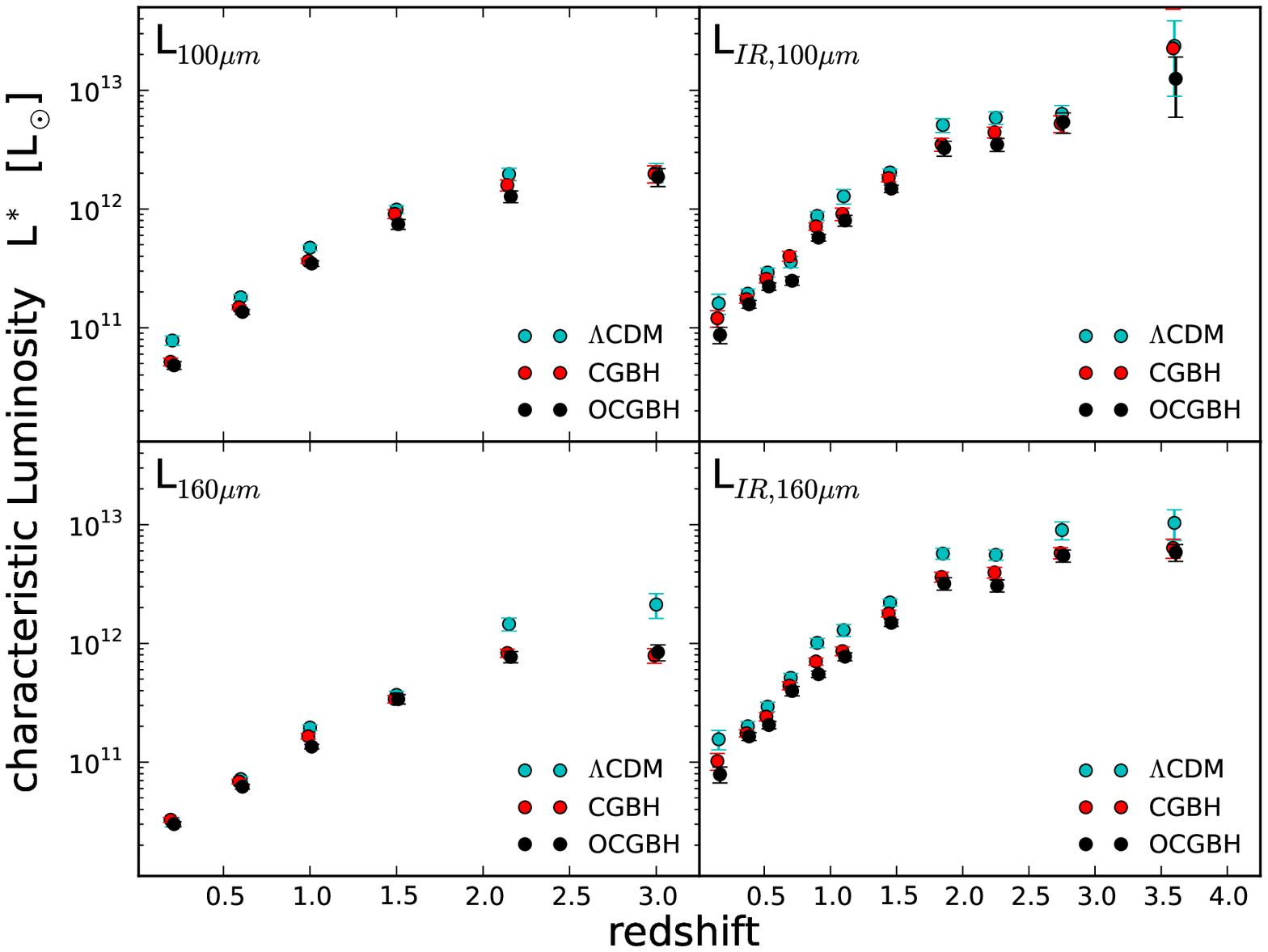}\\
\includegraphics[trim=0mm 0mm 0mm 0mm, clip, width=9cm]{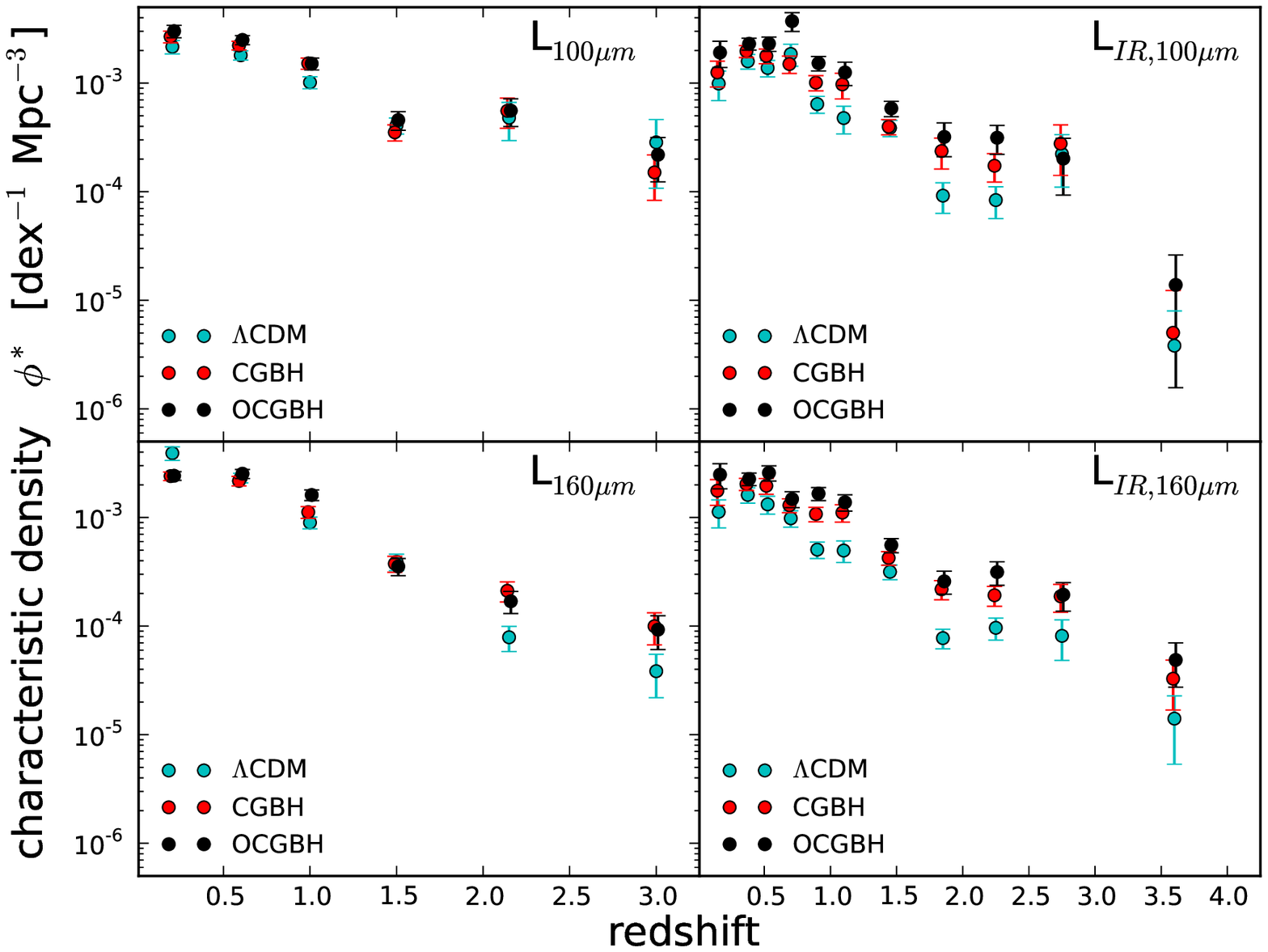}
\end{tabular}
\caption{{\em Upper panel:} Redshift evolution of the characteristic luminosity $L^{\ssty \ast}$ on the four datasets of the present work. {\em Lower panel:} Redshift evolution of the characteristic luminosity $\phi^{\ssty \ast}$
on the same datasets.
\label{evolfig}}
\end{figure}

We proceed to investigate the robustness with respect to the underlying cosmology of
the redshift evolution of the other two Schechter parameters: the characteristic luminosity,
$L^{\ssty \ast}$, and number density, $\phi^{\ssty \ast}$. { Figure \ref{evolfig} present the redshift evolution of theses parameters, which we model by means of
the simple relations:}
\begin{eqnarray}
L^{\ssty \ast}(z) \propto 10^{(1+z)  A} \\
\phi^{\ssty \ast}(z) \propto 10^{(1+z)  B}.
\end{eqnarray}
{ We use a least-squares technique to fit such evolution functions to their corresponding
Schechter parameter results \citep{1995TDAPR.122..107R}}.
Table \ref{evolt} lists the best { fit} values for the evolution parameters $A$ and $B$ in
the different datasets / cosmologies.

{ The listed uncertainties for the evolution parameters are the formal 1-$\sigma$
values obtained from the square root of the corresponding diagonal element of the
covariance matrix of the fit.}
We find no evidence for a significantly different evolution of
either $L^{\ssty \ast}$ or $\phi^{\ssty \ast}$ in the void models considered. The
monochromatic luminosities, specially the number density of sources in the rest-frame
160 $\mu$m, show some mild evidence of being affected by the geometrical effect discussed
above, but the evolution parameters in the total IR are remarkably similar. We also note
that assuming an open or flat CGBH model makes no significant difference to such parameters.
It seems that those are more strongly affected by the intrinsic evolution of the sources,
and the secular processes and merging history of galaxy formation, then the expansion
rate of the Universe.

{ Physically speaking, in terms of tracing the redshift evolution of different galaxy
populations using the FIR data in the present work, the marginally significant difference
in the faint-end slopes, together with the evolution parameters for the characteristic
number densities and luminosities, can be understood as follows: assertions about the
number density of FIR low-luminosity galaxies, broadly related to 
populations that
are poor in dust content, are still systematically affected by model-dependent corrections due to survey flux limits in the construction of the LF. That is, there might be less of those galaxies in the local Universe ($z \approx 0.3$), then what we expect based on the underlying standard model.
On the other hand, evolution of the
FIR high-luminosity end, broadly related to 
populations with high dust
content, is well constrained by the flux limits of the PEP survey, where the underlying
cosmological model is concerned.}

\section{Conclusions}
\label{conclusions}

In this work we have computed the far-IR luminosity functions for sources in the PEP
survey, observed at the Herschel/PACS 100 and 160 $\mu$m bands. We computed both
monochromatic and total IR luminosities assuming both the $\Lambda$CDM standard and GBH
void cosmological models, with aims to assess how robust the luminosity functions are
under a change of observationally constrained cosmologies.

We conclude that the current observational
constraints imposed on any cosmological model by the combined set of SNe + CMB + BAO
results are enough to yield robust estimates for the evolution of FIR characteristic
luminosities $L^{\ssty \ast}$ and number densities, $\phi^{\ssty \ast}$.

We find, however, that estimations of the faint-end slope of the LF are still
significantly dependent on the underlying cosmological model assumed, despite the
before mentioned observational constraints. That is, if there is indeed an underdense
region around the Galaxy, as predicted by the GBH models, causing the effective
metric of the Universe at Gpc scale to be better fit by an LTB line element, then
assuming the spatial homogeneous $\Lambda$CDM model in the computation of the LF would
yield a over-estimated number density of faint galaxies, at least at lower redshifts
(up to $z \approx 0.4$).

To answer the original question posed: the characteristic number density and the
characteristic luminosity parameters of the FIR luminosity functions derived here
are made robust by the present constraints on the cosmological model.
The faint end slope, however, still show significant differences among the
cosmologies studied here.

{ We show that those differences are caused mainly by slight discrepancies in the
luminosity distance - redshift relation, still allowed by the observations. The
1/V$_{\ssty max}$ methodology studied here is a necessary way to
compute the LF using a flux limited survey
like PEP.
Such methodology, as we show, is not biased
by the kind of under-dense regions proposed by the alternative cosmologies studied
here. On the other hand, the necessary volume corrections intrinsic in the method are still dependent enough on the underlying assumptions about the geometry and expansion rate of the Universe at
Gpc scale, to yield significant ($\approx$ 3-$\sigma$) discrepancies in their results.
In other words, the ``systematic'' dispersion in the values of the low luminosity LF points,
caused by the (arguably still) remaining degree of freedom in the choice of the underlying cosmological
model, combined with the current flux limits, is still significantly larger then the
statistic uncertainty assumed in the computation of the error bars of those points, causing the differences in the LF values to be larger then the combination of their computed uncertainties.

Surveys with lower flux limits would allow lower FIR-luminosity sources to be fully
accounted for, reducing the marginally significant dependency of the FIR
LF on the cosmological model still detected here.}

%
%

\section*{Aknowledgements}
AI is grateful to Takamitsu Miyagi and Karina Caputi, for the helpful tips on the building
of luminosity functions, and Alan Heavens, for the insightful discussions on bayesian and
Monte Carlo methods. Some Markov chain routines used at intermediate stages of this
work were derived as a spin-off of a workshop at the Cape Town International Cosmology
School, organized by the African Institute for Mathematical Sciences, between 15th and
28th of January, 2012. This work is jointly supported by Brazil's CAPES and ESO studentships.


\clearpage
\begin{table*}
\begin{tiny}
\caption[Luminosity Function in the 100 $\mu$m band, $\Lambda$CDM model]{Rest frame 100 $\mu$m 1/V$_{\ssty max}$ Luminosity Function assuming the $\Lambda$CDM cosmological model. Units are dex$^{-1}$.Mpc$^{-3}$.
\label{vmt100LCDM}}
\centering
\begin{tabular}{lllllll}
\hline \hline
\multicolumn{1}{c}{} &\multicolumn{ 6}{c}{Average redshift} \\
\multicolumn{1}{c}{Luminosity [L$_{\ssty \odot}$]}
 &\multicolumn{1}{c}{     0.2}  &\multicolumn{1}{c}{     0.6}
 &\multicolumn{1}{c}{     1.0}  &\multicolumn{1}{c}{     1.5}
 &\multicolumn{1}{c}{     2.1}  &\multicolumn{1}{c}{     3.0}
\\
\hline
\multicolumn{1}{c}{   5.0E+07}  &$(4.6 \pm 3.3) \times 10^{-2}$ &  &  &  &  & 
\\
\multicolumn{1}{c}{   1.3E+08}  &$(4.9 \pm 3.6) \times 10^{-2}$ &  &  &  &  & 
\\
\multicolumn{1}{c}{   3.2E+08}  &$(1.3 \pm 0.6) \times 10^{-2}$ &  &  &  &  & 
\\
\multicolumn{1}{c}{   7.9E+08}  &$(2.1 \pm 0.6) \times 10^{-2}$ &  &  &  &  & 
\\
\multicolumn{1}{c}{   2.0E+09}  &$(1.1 \pm 0.2) \times 10^{-2}$ &  &  &  &  & 
\\
\multicolumn{1}{c}{   5.0E+09}  &$(6.4 \pm 0.7) \times 10^{-3}$
 &$(2.9 \pm 2.9) \times 10^{-3}$ &  &  &  & 
\\
\multicolumn{1}{c}{   1.3E+10}  &$(4.3 \pm 0.4) \times 10^{-3}$
 &$(3.7 \pm 1.5) \times 10^{-3}$  &$(1.8 \pm 1.8) \times 10^{-4}$ &  &  & 
\\
\multicolumn{1}{c}{   3.2E+10}  &$(2.01 \pm 0.09) \times 10^{-3}$
 &$(3.4 \pm 0.4) \times 10^{-3}$  &$(4.4 \pm 1.8) \times 10^{-4}$
 &$(7.9 \pm 7.6) \times 10^{-6}$ &  & 
\\
\multicolumn{1}{c}{   7.9E+10}  &$(8.3 \pm 0.5) \times 10^{-4}$
 &$(2.3 \pm 0.2) \times 10^{-3}$  &$(1.3 \pm 0.2) \times 10^{-3}$
 &$(6.5 \pm 3.8) \times 10^{-5}$ &  & 
\\
\multicolumn{1}{c}{   2.0E+11}  &$(1.0 \pm 0.2) \times 10^{-4}$
 &$(5.2 \pm 0.2) \times 10^{-4}$  &$(1.3 \pm 0.2) \times 10^{-3}$
 &$(8.0 \pm 1.9) \times 10^{-4}$  &$(8.1 \pm 3.9) \times 10^{-5}$ & 
\\
\multicolumn{1}{c}{   5.0E+11}  &$(1.7 \pm 0.7) \times 10^{-5}$
 &$(8.2 \pm 0.7) \times 10^{-5}$  &$(3.8 \pm 0.4) \times 10^{-4}$
 &$(3.9 \pm 0.6) \times 10^{-4}$  &$(6.9 \pm 3.8) \times 10^{-4}$
 &$(1.2 \pm 0.6) \times 10^{-4}$
\\
\multicolumn{1}{c}{   1.3E+12} &   &$(6.8 \pm 2.0) \times 10^{-6}$
 &$(4.4 \pm 0.4) \times 10^{-5}$  &$(9.3 \pm 0.9) \times 10^{-5}$
 &$(3.1 \pm 1.2) \times 10^{-4}$  &$(5.2 \pm 4.1) \times 10^{-4}$
\\
\multicolumn{1}{c}{   3.2E+12} &  &   &$(1.9 \pm 0.8) \times 10^{-6}$
 &$(1.1 \pm 0.2) \times 10^{-5}$  &$(7.7 \pm 2.6) \times 10^{-5}$
 &$(4.8 \pm 1.5) \times 10^{-5}$
\\
\multicolumn{1}{c}{   7.9E+12} &  &  &   &$(3.7 \pm 2.6) \times 10^{-7}$
 &$(4.8 \pm 1.3) \times 10^{-6}$  &$(3.3 \pm 1.0) \times 10^{-6}$
\\
\multicolumn{1}{c}{   2.0E+13} &  &  &  &  &   &$(1.7 \pm 1.2) \times 10^{-6}$
\\
\hline
\end{tabular}
\end{tiny}
\end{table*}

\begin{table*}
\begin{tiny}
\caption[Luminosity Function in the 100 $\mu$m band, CGBH model]{Rest frame 100 $\mu$m 1/V$_{\ssty max}$ Luminosity Function assuming the CGBH cosmological model. Units are dex$^{-1}$.Mpc$^{-3}$.
\label{vmt100cGBH}}
\centering
\begin{tabular}{lllllll}
\hline \hline
\multicolumn{1}{c}{} &\multicolumn{ 6}{c}{Average redshift} \\
\multicolumn{1}{c}{Luminosity [L$_{\ssty \odot}$]}
 &\multicolumn{1}{c}{     0.2}  &\multicolumn{1}{c}{     0.6}
 &\multicolumn{1}{c}{     1.0}  &\multicolumn{1}{c}{     1.5}
 &\multicolumn{1}{c}{     2.1}  &\multicolumn{1}{c}{     3.0}
\\
\hline
\multicolumn{1}{c}{   5.0E+07}  &$(2.8 \pm 1.6) \times 10^{-3}$ &  &  &  &  & 
\\
\multicolumn{1}{c}{   1.3E+08}  &$(3.6 \pm 3.1) \times 10^{-3}$ &  &  &  &  & 
\\
\multicolumn{1}{c}{   3.2E+08}  &$(1.8 \pm 0.8) \times 10^{-3}$ &  &  &  &  & 
\\
\multicolumn{1}{c}{   7.9E+08}  &$(3.8 \pm 1.3) \times 10^{-3}$ &  &  &  &  & 
\\
\multicolumn{1}{c}{   2.0E+09}  &$(3.6 \pm 0.9) \times 10^{-3}$ &  &  &  &  & 
\\
\multicolumn{1}{c}{   5.0E+09}  &$(2.7 \pm 0.4) \times 10^{-3}$
 &$(2.2 \pm 2.2) \times 10^{-4}$ &  &  &  & 
\\
\multicolumn{1}{c}{   1.3E+10}  &$(2.3 \pm 0.2) \times 10^{-3}$
 &$(8.1 \pm 2.5) \times 10^{-4}$  &$(3.8 \pm 3.8) \times 10^{-5}$ &  &  & 
\\
\multicolumn{1}{c}{   3.2E+10}  &$(1.4 \pm 0.1) \times 10^{-3}$
 &$(2.6 \pm 0.4) \times 10^{-3}$  &$(4.3 \pm 2.6) \times 10^{-4}$
 &$(7.6 \pm 7.1) \times 10^{-6}$ &  & 
\\
\multicolumn{1}{c}{   7.9E+10}  &$(5.6 \pm 0.4) \times 10^{-4}$
 &$(1.8 \pm 0.2) \times 10^{-3}$  &$(1.1 \pm 0.2) \times 10^{-3}$
 &$(1.2 \pm 0.8) \times 10^{-4}$ &  & 
\\
\multicolumn{1}{c}{   2.0E+11}  &$(5.7 \pm 0.9) \times 10^{-5}$
 &$(5.1 \pm 0.3) \times 10^{-4}$  &$(1.1 \pm 0.1) \times 10^{-3}$
 &$(7.2 \pm 2.1) \times 10^{-4}$  &$(1.5 \pm 0.6) \times 10^{-4}$
 &$(1.6 \pm 1.6) \times 10^{-5}$
\\
\multicolumn{1}{c}{   5.0E+11}  &$(1.0 \pm 0.4) \times 10^{-5}$
 &$(8.2 \pm 0.7) \times 10^{-5}$  &$(3.5 \pm 0.4) \times 10^{-4}$
 &$(3.6 \pm 0.5) \times 10^{-4}$  &$(5.8 \pm 2.1) \times 10^{-4}$
 &$(7.5 \pm 5.3) \times 10^{-4}$
\\
\multicolumn{1}{c}{   1.3E+12} &   &$(5.9 \pm 1.7) \times 10^{-6}$
 &$(4.7 \pm 0.4) \times 10^{-5}$  &$(7.8 \pm 0.6) \times 10^{-5}$
 &$(2.2 \pm 0.8) \times 10^{-4}$  &$(7.2 \pm 2.7) \times 10^{-5}$
\\
\multicolumn{1}{c}{   3.2E+12} &  &   &$(2.2 \pm 0.9) \times 10^{-6}$
 &$(1.2 \pm 0.2) \times 10^{-5}$  &$(7.0 \pm 2.1) \times 10^{-5}$
 &$(5.3 \pm 2.2) \times 10^{-5}$
\\
\multicolumn{1}{c}{   7.9E+12} &  &  &  &   &$(3.8 \pm 1.4) \times 10^{-6}$
 &$(2.4 \pm 1.2) \times 10^{-6}$
\\
\multicolumn{1}{c}{   2.0E+13} &  &  &  &  &   &$(5.4 \pm 3.9) \times 10^{-7}$
\\
\hline
\end{tabular}
\end{tiny}
\end{table*}

\begin{table*}
\begin{tiny}
\caption[Luminosity Function in the 100 $\mu$m band, OCGBH model]{Rest frame 100 $\mu$m 1/V$_{\ssty max}$ Luminosity Function assuming the OCGBH cosmological model. Units are dex$^{-1}$.Mpc$^{-3}$.
\label{vmt100oGBH}}
\centering
\begin{tabular}{lllllll}
\hline \hline
\multicolumn{1}{c}{} &\multicolumn{ 6}{c}{Average redshift} \\
\multicolumn{1}{c}{Luminosity [L$_{\ssty \odot}$]}
 &\multicolumn{1}{c}{     0.2}  &\multicolumn{1}{c}{     0.6}
 &\multicolumn{1}{c}{     1.0}  &\multicolumn{1}{c}{     1.5}
 &\multicolumn{1}{c}{     2.1}  &\multicolumn{1}{c}{     3.0}
\\
\hline
\multicolumn{1}{c}{   5.0E+07}  &$(2.2 \pm 1.6) \times 10^{-3}$ &  &  &  &  & 
\\
\multicolumn{1}{c}{   1.3E+08}  &$(4.8 \pm 3.8) \times 10^{-3}$ &  &  &  &  & 
\\
\multicolumn{1}{c}{   3.2E+08}  &$(1.9 \pm 1.0) \times 10^{-3}$ &  &  &  &  & 
\\
\multicolumn{1}{c}{   7.9E+08}  &$(6.0 \pm 1.8) \times 10^{-3}$ &  &  &  &  & 
\\
\multicolumn{1}{c}{   2.0E+09}  &$(3.2 \pm 0.5) \times 10^{-3}$ &  &  &  &  & 
\\
\multicolumn{1}{c}{   5.0E+09}  &$(3.1 \pm 0.4) \times 10^{-3}$
 &$(3.1 \pm 2.6) \times 10^{-4}$ &  &  &  & 
\\
\multicolumn{1}{c}{   1.3E+10}  &$(2.7 \pm 0.2) \times 10^{-3}$
 &$(1.6 \pm 0.4) \times 10^{-3}$  &$(4.7 \pm 4.7) \times 10^{-5}$ &  &  & 
\\
\multicolumn{1}{c}{   3.2E+10}  &$(1.5 \pm 0.2) \times 10^{-3}$
 &$(2.8 \pm 0.4) \times 10^{-3}$  &$(7.1 \pm 3.4) \times 10^{-4}$
 &$(9.7 \pm 9.1) \times 10^{-6}$ &  & 
\\
\multicolumn{1}{c}{   7.9E+10}  &$(5.4 \pm 0.4) \times 10^{-4}$
 &$(2.0 \pm 0.2) \times 10^{-3}$  &$(1.5 \pm 0.3) \times 10^{-3}$
 &$(2.0 \pm 1.1) \times 10^{-4}$  &$(4.0 \pm 2.4) \times 10^{-5}$ & 
\\
\multicolumn{1}{c}{   2.0E+11}  &$(5.4 \pm 0.9) \times 10^{-5}$
 &$(5.1 \pm 0.3) \times 10^{-4}$  &$(1.3 \pm 0.2) \times 10^{-3}$
 &$(1.0 \pm 0.3) \times 10^{-3}$  &$(5.6 \pm 2.5) \times 10^{-4}$
 &$(9.1 \pm 6.0) \times 10^{-5}$
\\
\multicolumn{1}{c}{   5.0E+11}  &$(1.1 \pm 0.5) \times 10^{-5}$
 &$(7.2 \pm 0.7) \times 10^{-5}$  &$(2.9 \pm 0.3) \times 10^{-4}$
 &$(3.3 \pm 0.5) \times 10^{-4}$  &$(3.7 \pm 0.9) \times 10^{-4}$
 &$(8.5 \pm 6.5) \times 10^{-4}$
\\
\multicolumn{1}{c}{   1.3E+12} &   &$(6.0 \pm 1.9) \times 10^{-6}$
 &$(4.3 \pm 0.5) \times 10^{-5}$  &$(7.5 \pm 0.6) \times 10^{-5}$
 &$(2.7 \pm 0.9) \times 10^{-4}$  &$(1.2 \pm 0.4) \times 10^{-4}$
\\
\multicolumn{1}{c}{   3.2E+12} &  &   &$(1.4 \pm 0.8) \times 10^{-6}$
 &$(8.8 \pm 1.7) \times 10^{-6}$  &$(4.2 \pm 1.3) \times 10^{-5}$
 &$(3.3 \pm 1.6) \times 10^{-5}$
\\
\multicolumn{1}{c}{   7.9E+12} &  &  &  &   &$(1.1 \pm 0.7) \times 10^{-6}$
 &$(3.2 \pm 1.5) \times 10^{-6}$
\\
\multicolumn{1}{c}{   2.0E+13} &  &  &  &  &  & 
\\
\hline
\end{tabular}
\end{tiny}
\end{table*}

\clearpage
\begin{table*}
\begin{tiny}
\caption[Luminosity Function in the 160 $\mu$m band, $\Lambda$CDM model]{Rest frame 160 $\mu$m 1/V$_{\ssty max}$ Luminosity Function assuming the $\Lambda$CDM cosmological model. Units are dex$^{-1}$.Mpc$^{-3}$.
\label{vmt160LCDM}}
\centering
\begin{tabular}{lllllll}
\hline \hline
\multicolumn{1}{c}{} &\multicolumn{ 6}{c}{Average redshift} \\
\multicolumn{1}{c}{Luminosity [L$_{\ssty \odot}$]}
 &\multicolumn{1}{c}{     0.2}  &\multicolumn{1}{c}{     0.6}
 &\multicolumn{1}{c}{     1.0}  &\multicolumn{1}{c}{     1.5}
 &\multicolumn{1}{c}{     2.1}  &\multicolumn{1}{c}{     3.0}
\\
\hline
\multicolumn{1}{c}{   5.0E+07}  &$(4.9 \pm 2.8) \times 10^{-2}$ &  &  &  &  & 
\\
\multicolumn{1}{c}{   1.3E+08}  &$(6.6 \pm 4.3) \times 10^{-2}$ &  &  &  &  & 
\\
\multicolumn{1}{c}{   3.2E+08}  &$(6.7 \pm 3.2) \times 10^{-3}$ &  &  &  &  & 
\\
\multicolumn{1}{c}{   7.9E+08}  &$(1.7 \pm 0.5) \times 10^{-2}$ &  &  &  &  & 
\\
\multicolumn{1}{c}{   2.0E+09}  &$(1.2 \pm 0.2) \times 10^{-2}$ &  &  &  &  & 
\\
\multicolumn{1}{c}{   5.0E+09}  &$(5.2 \pm 0.6) \times 10^{-3}$
 &$(6.8 \pm 3.3) \times 10^{-3}$  &$(5.4 \pm 5.4) \times 10^{-5}$ &  &  & 
\\
\multicolumn{1}{c}{   1.3E+10}  &$(3.0 \pm 0.3) \times 10^{-3}$
 &$(2.7 \pm 0.4) \times 10^{-3}$  &$(1.8 \pm 0.8) \times 10^{-4}$
 &$(1.8 \pm 1.3) \times 10^{-5}$  &$(1.5 \pm 1.5) \times 10^{-5}$ & 
\\
\multicolumn{1}{c}{   3.2E+10}  &$(1.45 \pm 0.07) \times 10^{-3}$
 &$(2.1 \pm 0.2) \times 10^{-3}$  &$(8.4 \pm 1.5) \times 10^{-4}$
 &$(5.9 \pm 2.3) \times 10^{-4}$  &$(3.5 \pm 1.8) \times 10^{-5}$
 &$(2.2 \pm 1.8) \times 10^{-5}$
\\
\multicolumn{1}{c}{   7.9E+10}  &$(2.3 \pm 0.3) \times 10^{-4}$
 &$(7.0 \pm 0.4) \times 10^{-4}$  &$(1.1 \pm 0.1) \times 10^{-3}$
 &$(9.6 \pm 3.7) \times 10^{-4}$  &$(1.2 \pm 0.3) \times 10^{-4}$
 &$(3.5 \pm 1.3) \times 10^{-5}$
\\
\multicolumn{1}{c}{   2.0E+11}  &$(2.6 \pm 0.9) \times 10^{-5}$
 &$(1.13 \pm 0.08) \times 10^{-4}$  &$(3.4 \pm 0.4) \times 10^{-4}$
 &$(2.8 \pm 0.4) \times 10^{-4}$  &$(2.0 \pm 0.5) \times 10^{-4}$
 &$(1.2 \pm 0.7) \times 10^{-4}$
\\
\multicolumn{1}{c}{   5.0E+11} &   &$(3.0 \pm 1.3) \times 10^{-6}$
 &$(5.0 \pm 0.6) \times 10^{-5}$  &$(8.6 \pm 1.1) \times 10^{-5}$
 &$(1.2 \pm 0.5) \times 10^{-4}$  &$(6.1 \pm 2.2) \times 10^{-5}$
\\
\multicolumn{1}{c}{   1.3E+12} &  &   &$(2.1 \pm 0.9) \times 10^{-6}$
 &$(9.7 \pm 1.6) \times 10^{-6}$  &$(6.3 \pm 1.9) \times 10^{-5}$
 &$(2.2 \pm 0.6) \times 10^{-5}$
\\
\multicolumn{1}{c}{   3.2E+12} &  &  &   &$(5.8 \pm 3.4) \times 10^{-7}$
 &$(6.0 \pm 1.6) \times 10^{-6}$  &$(5.5 \pm 3.0) \times 10^{-6}$
\\
\multicolumn{1}{c}{   7.9E+12} &  &  &  &   &$(7.8 \pm 3.5) \times 10^{-7}$
 &$(1.1 \pm 0.5) \times 10^{-6}$
\\
\multicolumn{1}{c}{   2.0E+13} &  &  &  &  &  & 
\\
\hline
\end{tabular}
\end{tiny}
\end{table*}

\begin{table*}
\begin{tiny}
\caption[Luminosity Function in the 160 $\mu$m band, CGBH model]{Rest frame 160 $\mu$m 1/V$_{\ssty max}$ Luminosity Function assuming the CGBH cosmological model. Units are dex$^{-1}$.Mpc$^{-3}$.
\label{vmt160cGBH}}
\centering
\begin{tabular}{lllllll}
\hline \hline
\multicolumn{1}{c}{} &\multicolumn{ 6}{c}{Average redshift} \\
\multicolumn{1}{c}{Luminosity [L$_{\ssty \odot}$]}
 &\multicolumn{1}{c}{     0.2}  &\multicolumn{1}{c}{     0.6}
 &\multicolumn{1}{c}{     1.0}  &\multicolumn{1}{c}{     1.5}
 &\multicolumn{1}{c}{     2.1}  &\multicolumn{1}{c}{     3.0}
\\
\hline
\multicolumn{1}{c}{   5.0E+07}  &$(2.3 \pm 1.4) \times 10^{-3}$ &  &  &  &  & 
\\
\multicolumn{1}{c}{   1.3E+08}  &$(3.5 \pm 2.6) \times 10^{-3}$ &  &  &  &  & 
\\
\multicolumn{1}{c}{   3.2E+08}  &$(7.8 \pm 2.8) \times 10^{-4}$ &  &  &  &  & 
\\
\multicolumn{1}{c}{   7.9E+08}  &$(2.6 \pm 1.0) \times 10^{-3}$ &  &  &  &  & 
\\
\multicolumn{1}{c}{   2.0E+09}  &$(5.3 \pm 1.4) \times 10^{-3}$ &  &  &  &  & 
\\
\multicolumn{1}{c}{   5.0E+09}  &$(2.7 \pm 0.4) \times 10^{-3}$
 &$(4.4 \pm 1.8) \times 10^{-4}$  &$(4.5 \pm 4.5) \times 10^{-5}$ &  &  & 
\\
\multicolumn{1}{c}{   1.3E+10}  &$(2.4 \pm 0.2) \times 10^{-3}$
 &$(2.3 \pm 0.3) \times 10^{-3}$  &$(2.9 \pm 1.4) \times 10^{-4}$
 &$(1.5 \pm 1.1) \times 10^{-5}$  &$(2.0 \pm 2.0) \times 10^{-5}$ & 
\\
\multicolumn{1}{c}{   3.2E+10}  &$(1.0 \pm 0.1) \times 10^{-3}$
 &$(1.6 \pm 0.2) \times 10^{-3}$  &$(7.2 \pm 1.2) \times 10^{-4}$
 &$(3.5 \pm 1.2) \times 10^{-4}$  &$(7.1 \pm 2.9) \times 10^{-5}$
 &$(5.3 \pm 2.8) \times 10^{-5}$
\\
\multicolumn{1}{c}{   7.9E+10}  &$(1.8 \pm 0.2) \times 10^{-4}$
 &$(6.5 \pm 0.4) \times 10^{-4}$  &$(1.1 \pm 0.1) \times 10^{-3}$
 &$(5.7 \pm 1.4) \times 10^{-4}$  &$(1.5 \pm 0.4) \times 10^{-4}$
 &$(2.8 \pm 1.2) \times 10^{-5}$
\\
\multicolumn{1}{c}{   2.0E+11}  &$(1.6 \pm 0.5) \times 10^{-5}$
 &$(1.18 \pm 0.08) \times 10^{-4}$  &$(3.3 \pm 0.4) \times 10^{-4}$
 &$(2.3 \pm 0.3) \times 10^{-4}$  &$(2.9 \pm 0.7) \times 10^{-4}$
 &$(1.0 \pm 0.4) \times 10^{-4}$
\\
\multicolumn{1}{c}{   5.0E+11} &   &$(4.1 \pm 1.5) \times 10^{-6}$
 &$(5.1 \pm 0.6) \times 10^{-5}$  &$(7.2 \pm 1.0) \times 10^{-5}$
 &$(1.2 \pm 0.6) \times 10^{-4}$  &$(5.6 \pm 1.7) \times 10^{-5}$
\\
\multicolumn{1}{c}{   1.3E+12} &  &   &$(2.4 \pm 1.0) \times 10^{-6}$
 &$(1.0 \pm 0.2) \times 10^{-5}$  &$(7.8 \pm 2.6) \times 10^{-5}$
 &$(1.6 \pm 0.5) \times 10^{-5}$
\\
\multicolumn{1}{c}{   3.2E+12} &  &  &   &$(2.9 \pm 2.9) \times 10^{-7}$
 &$(4.3 \pm 1.1) \times 10^{-6}$  &$(2.2 \pm 0.8) \times 10^{-6}$
\\
\multicolumn{1}{c}{   7.9E+12} &  &  &  &   &$(8.8 \pm 4.4) \times 10^{-7}$
 &$(8.1 \pm 4.1) \times 10^{-7}$
\\
\multicolumn{1}{c}{   2.0E+13} &  &  &  &  &  & 
\\
\hline
\end{tabular}
\end{tiny}
\end{table*}

\begin{table*}
\begin{tiny}
\caption[Luminosity Function in the 160 $\mu$m band, OCGBH model]{Rest frame 160 $\mu$m 1/V$_{\ssty max}$ Luminosity Function assuming the OCGBH cosmological model. Units are dex$^{-1}$.Mpc$^{-3}$.
\label{vmt160oGBH}}
\centering
\begin{tabular}{lllllll}
\hline \hline
\multicolumn{1}{c}{} &\multicolumn{ 6}{c}{Average redshift} \\
\multicolumn{1}{c}{Luminosity [L$_{\ssty \odot}$]}
 &\multicolumn{1}{c}{     0.2}  &\multicolumn{1}{c}{     0.6}
 &\multicolumn{1}{c}{     1.0}  &\multicolumn{1}{c}{     1.5}
 &\multicolumn{1}{c}{     2.1}  &\multicolumn{1}{c}{     3.0}
\\
\hline
\multicolumn{1}{c}{   5.0E+07}  &$(2.9 \pm 1.8) \times 10^{-3}$ &  &  &  &  & 
\\
\multicolumn{1}{c}{   1.3E+08}  &$(4.7 \pm 3.1) \times 10^{-3}$ &  &  &  &  & 
\\
\multicolumn{1}{c}{   3.2E+08}  &$(5.3 \pm 2.4) \times 10^{-4}$ &  &  &  &  & 
\\
\multicolumn{1}{c}{   7.9E+08}  &$(3.7 \pm 1.2) \times 10^{-3}$ &  &  &  &  & 
\\
\multicolumn{1}{c}{   2.0E+09}  &$(6.3 \pm 1.6) \times 10^{-3}$ &  &  &  &  & 
\\
\multicolumn{1}{c}{   5.0E+09}  &$(3.0 \pm 0.4) \times 10^{-3}$
 &$(9.7 \pm 2.8) \times 10^{-4}$  &$(2.0 \pm 1.5) \times 10^{-4}$ &  &  & 
\\
\multicolumn{1}{c}{   1.3E+10}  &$(2.6 \pm 0.2) \times 10^{-3}$
 &$(2.6 \pm 0.4) \times 10^{-3}$  &$(3.3 \pm 1.2) \times 10^{-4}$
 &$(1.9 \pm 1.4) \times 10^{-5}$  &$(2.5 \pm 2.5) \times 10^{-5}$ & 
\\
\multicolumn{1}{c}{   3.2E+10}  &$(1.08 \pm 0.06) \times 10^{-3}$
 &$(2.0 \pm 0.2) \times 10^{-3}$  &$(1.1 \pm 0.2) \times 10^{-3}$
 &$(6.2 \pm 1.8) \times 10^{-4}$  &$(1.0 \pm 0.4) \times 10^{-4}$
 &$(8.2 \pm 3.6) \times 10^{-5}$
\\
\multicolumn{1}{c}{   7.9E+10}  &$(1.3 \pm 0.2) \times 10^{-4}$
 &$(6.6 \pm 0.4) \times 10^{-4}$  &$(1.1 \pm 0.1) \times 10^{-3}$
 &$(5.7 \pm 1.5) \times 10^{-4}$  &$(2.2 \pm 0.5) \times 10^{-4}$
 &$(4.9 \pm 2.2) \times 10^{-5}$
\\
\multicolumn{1}{c}{   2.0E+11}  &$(1.5 \pm 0.5) \times 10^{-5}$
 &$(1.07 \pm 0.09) \times 10^{-4}$  &$(3.4 \pm 0.4) \times 10^{-4}$
 &$(3.0 \pm 0.4) \times 10^{-4}$  &$(4.1 \pm 1.1) \times 10^{-4}$
 &$(1.2 \pm 0.4) \times 10^{-4}$
\\
\multicolumn{1}{c}{   5.0E+11} &   &$(6.8 \pm 6.8) \times 10^{-7}$
 &$(4.0 \pm 0.5) \times 10^{-5}$  &$(7.0 \pm 0.7) \times 10^{-5}$
 &$(7.2 \pm 1.3) \times 10^{-5}$  &$(5.0 \pm 1.5) \times 10^{-5}$
\\
\multicolumn{1}{c}{   1.3E+12} &  &  &   &$(1.1 \pm 0.2) \times 10^{-5}$
 &$(9.5 \pm 3.2) \times 10^{-5}$  &$(1.7 \pm 0.6) \times 10^{-5}$
\\
\multicolumn{1}{c}{   3.2E+12} &  &  &  &   &$(2.9 \pm 1.0) \times 10^{-6}$
 &$(2.5 \pm 0.9) \times 10^{-6}$
\\
\multicolumn{1}{c}{   7.9E+12} &  &  &  &   &$(5.5 \pm 3.9) \times 10^{-7}$
 &$(7.8 \pm 4.5) \times 10^{-7}$
\\
\multicolumn{1}{c}{   2.0E+13} &  &  &  &  &  & 
\\
\hline
\end{tabular}
\end{tiny}
\end{table*}

\clearpage
\begin{sidewaystable}
\setlength{\tabcolsep}{1.25pt}
\begin{scriptsize}
\caption[Total IR Luminosity Function in the 100 $\mu$m band, $\Lambda$CDM model]{Rest frame total IR 1/V$_{\ssty max}$ Luminosity Function in the PACS 100 $\mu$m band, assuming the $\Lambda$CDM cosmological model. Units are dex$^{-1}$.Mpc$^{-3}$.
\label{vmtLIR100LCDM}}
\centering
\begin{tabular}{llllllllllll}
\hline \hline
\multicolumn{1}{c}{} &\multicolumn{ 11}{c}{Average redshift} \\
\multicolumn{1}{c}{Luminosity [L$_{\ssty \odot}$]}
 &\multicolumn{1}{c}{     0.2}  &\multicolumn{1}{c}{     0.4}
 &\multicolumn{1}{c}{     0.5}  &\multicolumn{1}{c}{     0.7}
 &\multicolumn{1}{c}{     0.9}  &\multicolumn{1}{c}{     1.1}
 &\multicolumn{1}{c}{     1.5}  &\multicolumn{1}{c}{     1.9}
 &\multicolumn{1}{c}{     2.2}  &\multicolumn{1}{c}{     2.8}
 &\multicolumn{1}{c}{     3.6}
\\
\hline
\multicolumn{1}{c}{   1.8E+08}  &$(3.7 \pm 2.8) \times 10^{-2}$ &  &  &  &  & 
&  &  &  &  & 
\\
\multicolumn{1}{c}{   5.6E+08}  &$(1.3 \pm 0.5) \times 10^{-2}$ &  &  &  &  & 
&  &  &  &  & 
\\
\multicolumn{1}{c}{   1.8E+09}  &$(1.9 \pm 0.5) \times 10^{-2}$ &  &  &  &  & 
&  &  &  &  & 
\\
\multicolumn{1}{c}{   5.6E+09}  &$(9.3 \pm 1.3) \times 10^{-3}$
 &$(1.0 \pm 0.8) \times 10^{-3}$ &  &  &  &  &  &  &  &  & 
\\
\multicolumn{1}{c}{   1.8E+10}  &$(3.7 \pm 0.3) \times 10^{-3}$
 &$(4.7 \pm 1.1) \times 10^{-3}$  &$(5.8 \pm 3.4) \times 10^{-4}$ &  &  & 
 &$(1.0 \pm 1.0) \times 10^{-6}$ &  &  &  & 
\\
\multicolumn{1}{c}{   5.6E+10}  &$(1.46 \pm 0.09) \times 10^{-3}$
 &$(2.8 \pm 0.2) \times 10^{-3}$  &$(2.9 \pm 0.5) \times 10^{-3}$
 &$(1.3 \pm 0.9) \times 10^{-2}$  &$(5.6 \pm 2.2) \times 10^{-4}$ &  &  &  &  & 
& 
\\
\multicolumn{1}{c}{   1.8E+11}  &$(2.8 \pm 0.4) \times 10^{-4}$
 &$(6.6 \pm 0.4) \times 10^{-4}$  &$(1.2 \pm 0.1) \times 10^{-3}$
 &$(2.1 \pm 0.2) \times 10^{-3}$  &$(1.3 \pm 0.2) \times 10^{-3}$
 &$(1.9 \pm 0.5) \times 10^{-3}$  &$(2.2 \pm 1.2) \times 10^{-4}$ &  &  &  & 
\\
\multicolumn{1}{c}{   5.6E+11}  &$(2.0 \pm 1.0) \times 10^{-5}$
 &$(4.6 \pm 1.1) \times 10^{-5}$  &$(1.2 \pm 0.1) \times 10^{-4}$
 &$(2.7 \pm 0.2) \times 10^{-4}$  &$(5.3 \pm 0.5) \times 10^{-4}$
 &$(6.6 \pm 1.1) \times 10^{-4}$  &$(6.7 \pm 1.0) \times 10^{-4}$
 &$(2.2 \pm 0.9) \times 10^{-4}$  &$(2.3 \pm 0.9) \times 10^{-4}$ &  & 
\\
\multicolumn{1}{c}{   1.8E+12}  &$(4.9 \pm 4.9) \times 10^{-6}$
 &$(2.4 \pm 2.4) \times 10^{-6}$  &$(2.9 \pm 2.1) \times 10^{-6}$
 &$(1.7 \pm 0.4) \times 10^{-5}$  &$(5.0 \pm 0.5) \times 10^{-5}$
 &$(8.9 \pm 0.7) \times 10^{-5}$  &$(1.8 \pm 0.2) \times 10^{-4}$
 &$(2.1 \pm 0.5) \times 10^{-4}$  &$(3.9 \pm 1.2) \times 10^{-4}$
 &$(4.6 \pm 3.3) \times 10^{-4}$ & 
\\
\multicolumn{1}{c}{   5.6E+12} &  &  &   &$(1.5 \pm 1.1) \times 10^{-6}$
 &$(1.7 \pm 1.0) \times 10^{-6}$  &$(6.1 \pm 1.7) \times 10^{-6}$
 &$(1.2 \pm 0.1) \times 10^{-5}$  &$(3.9 \pm 0.8) \times 10^{-5}$
 &$(6.8 \pm 1.4) \times 10^{-5}$  &$(9.8 \pm 2.7) \times 10^{-5}$
 &$(2.8 \pm 1.4) \times 10^{-5}$
\\
\multicolumn{1}{c}{   1.8E+13} &  &  &  &  &  & 
 &$(3.0 \pm 2.1) \times 10^{-7}$  &$(9.5 \pm 4.7) \times 10^{-7}$
 &$(1.7 \pm 0.5) \times 10^{-6}$  &$(6.6 \pm 1.6) \times 10^{-6}$
 &$(1.9 \pm 0.7) \times 10^{-6}$
\\
\hline
\end{tabular}
\end{scriptsize}
\end{sidewaystable}

\begin{sidewaystable}
\setlength{\tabcolsep}{1.25pt}
\begin{scriptsize}
\caption[Total IR Luminosity Function in the 160 $\mu$m band, $\Lambda$CDM model]{Rest frame total IR 1/V$_{\ssty max}$ Luminosity Function in the PACS 160 $\mu$m band, assuming the $\Lambda$CDM cosmological model. Units are dex$^{-1}$.Mpc$^{-3}$.
\label{vmtLIR160LCDM}}
\centering
\begin{tabular}{llllllllllll}
\hline \hline
\multicolumn{1}{c}{} &\multicolumn{ 11}{c}{Average redshift} \\
\multicolumn{1}{c}{Luminosity [L$_{\ssty \odot}$]}
 &\multicolumn{1}{c}{     0.2}  &\multicolumn{1}{c}{     0.4}
 &\multicolumn{1}{c}{     0.5}  &\multicolumn{1}{c}{     0.7}
 &\multicolumn{1}{c}{     0.9}  &\multicolumn{1}{c}{     1.1}
 &\multicolumn{1}{c}{     1.5}  &\multicolumn{1}{c}{     1.9}
 &\multicolumn{1}{c}{     2.2}  &\multicolumn{1}{c}{     2.8}
 &\multicolumn{1}{c}{     3.6}
\\
\hline
\multicolumn{1}{c}{   1.8E+08}  &$(4.9 \pm 3.5) \times 10^{-2}$ &  &  &  &  & 
&  &  &  &  & 
\\
\multicolumn{1}{c}{   5.6E+08}  &$(7.7 \pm 4.1) \times 10^{-3}$ &  &  &  &  & 
&  &  &  &  & 
\\
\multicolumn{1}{c}{   1.8E+09}  &$(1.6 \pm 0.4) \times 10^{-2}$ &  &  &  &  & 
&  &  &  &  & 
\\
\multicolumn{1}{c}{   5.6E+09}  &$(9.0 \pm 1.3) \times 10^{-3}$
 &$(1.8 \pm 1.8) \times 10^{-3}$ &  &  &  &  &  &  &  &  & 
\\
\multicolumn{1}{c}{   1.8E+10}  &$(3.6 \pm 0.3) \times 10^{-3}$
 &$(4.0 \pm 1.0) \times 10^{-3}$  &$(5.8 \pm 2.7) \times 10^{-4}$ &  &  &  &  & 
&  &  & 
\\
\multicolumn{1}{c}{   5.6E+10}  &$(1.51 \pm 0.09) \times 10^{-3}$
 &$(2.5 \pm 0.3) \times 10^{-3}$  &$(2.8 \pm 0.4) \times 10^{-3}$
 &$(4.2 \pm 1.3) \times 10^{-3}$  &$(1.1 \pm 0.8) \times 10^{-4}$ &  &  &  &  & 
& 
\\
\multicolumn{1}{c}{   1.8E+11}  &$(3.1 \pm 0.4) \times 10^{-4}$
 &$(7.3 \pm 0.5) \times 10^{-4}$  &$(1.0 \pm 0.1) \times 10^{-3}$
 &$(1.5 \pm 0.2) \times 10^{-3}$  &$(9.0 \pm 1.5) \times 10^{-4}$
 &$(1.4 \pm 0.3) \times 10^{-3}$  &$(8.5 \pm 3.4) \times 10^{-4}$ &  &  &  & 
\\
\multicolumn{1}{c}{   5.6E+11}  &$(2.1 \pm 1.0) \times 10^{-5}$
 &$(4.9 \pm 1.1) \times 10^{-5}$  &$(1.3 \pm 0.1) \times 10^{-4}$
 &$(2.9 \pm 0.2) \times 10^{-4}$  &$(6.1 \pm 0.6) \times 10^{-4}$
 &$(7.4 \pm 1.0) \times 10^{-4}$  &$(5.2 \pm 0.7) \times 10^{-4}$
 &$(1.6 \pm 0.5) \times 10^{-4}$  &$(1.6 \pm 0.7) \times 10^{-4}$ &  & 
\\
\multicolumn{1}{c}{   1.8E+12}  &$(5.4 \pm 5.4) \times 10^{-6}$
 &$(2.5 \pm 2.5) \times 10^{-6}$  &$(3.0 \pm 2.1) \times 10^{-6}$
 &$(1.7 \pm 0.4) \times 10^{-5}$  &$(5.7 \pm 0.6) \times 10^{-5}$
 &$(9.6 \pm 0.7) \times 10^{-5}$  &$(1.7 \pm 0.2) \times 10^{-4}$
 &$(1.3 \pm 0.3) \times 10^{-4}$  &$(2.9 \pm 0.5) \times 10^{-4}$
 &$(1.4 \pm 0.6) \times 10^{-4}$  &$(2.5 \pm 1.6) \times 10^{-5}$
\\
\multicolumn{1}{c}{   5.6E+12} &  &  &   &$(1.5 \pm 1.1) \times 10^{-6}$
 &$(1.7 \pm 1.0) \times 10^{-6}$  &$(6.3 \pm 1.7) \times 10^{-6}$
 &$(1.4 \pm 0.2) \times 10^{-5}$  &$(3.6 \pm 0.3) \times 10^{-5}$
 &$(4.4 \pm 0.6) \times 10^{-5}$  &$(7.7 \pm 1.8) \times 10^{-5}$
 &$(2.5 \pm 0.8) \times 10^{-5}$
\\
\multicolumn{1}{c}{   1.8E+13} &  &  &  &  &  & 
 &$(3.2 \pm 2.3) \times 10^{-7}$  &$(9.6 \pm 4.8) \times 10^{-7}$
 &$(1.6 \pm 0.5) \times 10^{-6}$  &$(7.1 \pm 1.1) \times 10^{-6}$
 &$(1.6 \pm 0.5) \times 10^{-6}$
\\
\hline
\end{tabular}
\end{scriptsize}
\end{sidewaystable}

\clearpage
\begin{sidewaystable}
\setlength{\tabcolsep}{1.25pt}
\begin{scriptsize}
\caption[Total IR Luminosity Function in the 100 $\mu$m band, CGBH model]{Rest frame total IR 1/V$_{\ssty max}$ Luminosity Function in the PACS 100 $\mu$m band, assuming the CGBH cosmological model. Units are dex$^{-1}$.Mpc$^{-3}$.
\label{vmtLIR100cGBH}}
\centering
\begin{tabular}{llllllllllll}
\hline \hline
\multicolumn{1}{c}{} &\multicolumn{ 11}{c}{Average redshift} \\
\multicolumn{1}{c}{Luminosity [L$_{\ssty \odot}$]}
 &\multicolumn{1}{c}{     0.2}  &\multicolumn{1}{c}{     0.4}
 &\multicolumn{1}{c}{     0.5}  &\multicolumn{1}{c}{     0.7}
 &\multicolumn{1}{c}{     0.9}  &\multicolumn{1}{c}{     1.1}
 &\multicolumn{1}{c}{     1.5}  &\multicolumn{1}{c}{     1.9}
 &\multicolumn{1}{c}{     2.2}  &\multicolumn{1}{c}{     2.8}
 &\multicolumn{1}{c}{     3.6}
\\
\hline
\multicolumn{1}{c}{   1.8E+08}  &$(3.3 \pm 3.1) \times 10^{-3}$ &  &  &  &  & 
&  &  &  &  & 
\\
\multicolumn{1}{c}{   5.6E+08}  &$(2.0 \pm 0.9) \times 10^{-3}$ &  &  &  &  & 
&  &  &  &  & 
\\
\multicolumn{1}{c}{   1.8E+09}  &$(5.6 \pm 1.4) \times 10^{-3}$ &  &  &  &  & 
&  &  &  &  & 
\\
\multicolumn{1}{c}{   5.6E+09}  &$(3.9 \pm 0.5) \times 10^{-3}$
 &$(6.0 \pm 4.4) \times 10^{-4}$ &  &  &  &  &  &  &  &  & 
\\
\multicolumn{1}{c}{   1.8E+10}  &$(2.3 \pm 0.2) \times 10^{-3}$
 &$(2.6 \pm 0.6) \times 10^{-3}$  &$(1.8 \pm 1.8) \times 10^{-4}$ &  &  & 
 &$(5.7 \pm 5.7) \times 10^{-7}$ &  &  &  & 
\\
\multicolumn{1}{c}{   5.6E+10}  &$(1.1 \pm 0.1) \times 10^{-3}$
 &$(2.2 \pm 0.2) \times 10^{-3}$  &$(2.4 \pm 0.4) \times 10^{-3}$
 &$(4.2 \pm 1.7) \times 10^{-3}$  &$(5.3 \pm 2.4) \times 10^{-4}$ &  &  &  &  & 
& 
\\
\multicolumn{1}{c}{   1.8E+11}  &$(2.2 \pm 0.3) \times 10^{-4}$
 &$(7.1 \pm 0.4) \times 10^{-4}$  &$(1.07 \pm 0.09) \times 10^{-3}$
 &$(1.9 \pm 0.2) \times 10^{-3}$  &$(1.2 \pm 0.2) \times 10^{-3}$
 &$(1.8 \pm 0.5) \times 10^{-3}$  &$(3.2 \pm 1.7) \times 10^{-4}$ &  &  &  & 
\\
\multicolumn{1}{c}{   5.6E+11}  &$(1.2 \pm 0.5) \times 10^{-5}$
 &$(4.8 \pm 1.0) \times 10^{-5}$  &$(1.5 \pm 0.1) \times 10^{-4}$
 &$(3.0 \pm 0.2) \times 10^{-4}$  &$(5.7 \pm 0.6) \times 10^{-4}$
 &$(6.8 \pm 1.2) \times 10^{-4}$  &$(6.1 \pm 0.9) \times 10^{-4}$
 &$(2.9 \pm 1.2) \times 10^{-4}$  &$(2.2 \pm 0.9) \times 10^{-4}$ &  & 
\\
\multicolumn{1}{c}{   1.8E+12}  &$(2.9 \pm 2.9) \times 10^{-6}$
 &$(5.1 \pm 3.6) \times 10^{-6}$  &$(4.1 \pm 2.3) \times 10^{-6}$
 &$(1.7 \pm 0.4) \times 10^{-5}$  &$(5.7 \pm 0.6) \times 10^{-5}$
 &$(1.02 \pm 0.09) \times 10^{-4}$  &$(1.4 \pm 0.1) \times 10^{-4}$
 &$(2.7 \pm 0.8) \times 10^{-4}$  &$(4.7 \pm 1.2) \times 10^{-4}$
 &$(6.8 \pm 4.2) \times 10^{-4}$  &$(9.1 \pm 9.1) \times 10^{-6}$
\\
\multicolumn{1}{c}{   5.6E+12} &  &  &   &$(1.6 \pm 1.1) \times 10^{-6}$
 &$(1.8 \pm 1.0) \times 10^{-6}$  &$(7.3 \pm 2.0) \times 10^{-6}$
 &$(1.2 \pm 0.2) \times 10^{-5}$  &$(4.4 \pm 1.1) \times 10^{-5}$
 &$(5.4 \pm 1.3) \times 10^{-5}$  &$(8.7 \pm 2.6) \times 10^{-5}$
 &$(3.0 \pm 1.6) \times 10^{-5}$
\\
\multicolumn{1}{c}{   1.8E+13} &  &  &  &  &  & 
 &$(3.8 \pm 2.7) \times 10^{-7}$  &$(6.5 \pm 4.6) \times 10^{-7}$
 &$(1.7 \pm 0.6) \times 10^{-6}$  &$(5.9 \pm 1.6) \times 10^{-6}$
 &$(2.5 \pm 0.9) \times 10^{-6}$
\\
\hline
\end{tabular}
\end{scriptsize}
\end{sidewaystable}

\begin{sidewaystable}
\setlength{\tabcolsep}{1.25pt}
\begin{scriptsize}
\caption[Total IR Luminosity Function in the 160 $\mu$m band, CGBH model]{Rest frame total IR 1/V$_{\ssty max}$ Luminosity Function in the PACS 160 $\mu$m band, assuming the CGBH cosmological model. Units are dex$^{-1}$.Mpc$^{-3}$.
\label{vmtLIR160cGBH}}
\centering
\begin{tabular}{llllllllllll}
\hline \hline
\multicolumn{1}{c}{} &\multicolumn{ 11}{c}{Average redshift} \\
\multicolumn{1}{c}{Luminosity [L$_{\ssty \odot}$]}
 &\multicolumn{1}{c}{     0.2}  &\multicolumn{1}{c}{     0.4}
 &\multicolumn{1}{c}{     0.5}  &\multicolumn{1}{c}{     0.7}
 &\multicolumn{1}{c}{     0.9}  &\multicolumn{1}{c}{     1.1}
 &\multicolumn{1}{c}{     1.5}  &\multicolumn{1}{c}{     1.9}
 &\multicolumn{1}{c}{     2.2}  &\multicolumn{1}{c}{     2.8}
 &\multicolumn{1}{c}{     3.6}
\\
\hline
\multicolumn{1}{c}{   1.8E+08}  &$(3.3 \pm 2.8) \times 10^{-3}$ &  &  &  &  & 
&  &  &  &  & 
\\
\multicolumn{1}{c}{   5.6E+08}  &$(1.7 \pm 0.8) \times 10^{-3}$ &  &  &  &  & 
&  &  &  &  & 
\\
\multicolumn{1}{c}{   1.8E+09}  &$(3.8 \pm 1.2) \times 10^{-3}$ &  &  &  &  & 
&  &  &  &  & 
\\
\multicolumn{1}{c}{   5.6E+09}  &$(5.0 \pm 1.1) \times 10^{-3}$
 &$(2.9 \pm 2.9) \times 10^{-4}$ &  &  &  &  &  &  &  &  & 
\\
\multicolumn{1}{c}{   1.8E+10}  &$(2.3 \pm 0.2) \times 10^{-3}$
 &$(3.1 \pm 0.8) \times 10^{-3}$  &$(4.7 \pm 2.4) \times 10^{-4}$ &  &  &  &  & 
&  &  & 
\\
\multicolumn{1}{c}{   5.6E+10}  &$(1.3 \pm 0.1) \times 10^{-3}$
 &$(1.7 \pm 0.2) \times 10^{-3}$  &$(2.5 \pm 0.4) \times 10^{-3}$
 &$(2.3 \pm 0.4) \times 10^{-3}$  &$(8.9 \pm 6.4) \times 10^{-5}$ &  &  &  &  & 
& 
\\
\multicolumn{1}{c}{   1.8E+11}  &$(2.4 \pm 0.3) \times 10^{-4}$
 &$(8.0 \pm 0.5) \times 10^{-4}$  &$(9.8 \pm 1.2) \times 10^{-4}$
 &$(1.3 \pm 0.2) \times 10^{-3}$  &$(9.4 \pm 1.9) \times 10^{-4}$
 &$(1.3 \pm 0.2) \times 10^{-3}$  &$(3.6 \pm 1.3) \times 10^{-4}$ &  &  &  & 
\\
\multicolumn{1}{c}{   5.6E+11}  &$(1.2 \pm 0.5) \times 10^{-5}$
 &$(5.1 \pm 1.1) \times 10^{-5}$  &$(1.5 \pm 0.2) \times 10^{-4}$
 &$(3.3 \pm 0.2) \times 10^{-4}$  &$(5.9 \pm 0.6) \times 10^{-4}$
 &$(7.6 \pm 1.1) \times 10^{-4}$  &$(5.3 \pm 0.7) \times 10^{-4}$
 &$(2.4 \pm 0.7) \times 10^{-4}$  &$(2.1 \pm 0.7) \times 10^{-4}$
 &$(1.9 \pm 1.9) \times 10^{-5}$ & 
\\
\multicolumn{1}{c}{   1.8E+12}  &$(2.9 \pm 2.9) \times 10^{-6}$
 &$(5.8 \pm 4.1) \times 10^{-6}$  &$(4.2 \pm 2.4) \times 10^{-6}$
 &$(1.8 \pm 0.4) \times 10^{-5}$  &$(6.5 \pm 0.7) \times 10^{-5}$
 &$(1.12 \pm 0.09) \times 10^{-4}$  &$(1.4 \pm 0.1) \times 10^{-4}$
 &$(1.5 \pm 0.4) \times 10^{-4}$  &$(3.3 \pm 0.6) \times 10^{-4}$
 &$(1.8 \pm 0.5) \times 10^{-4}$  &$(2.7 \pm 1.4) \times 10^{-5}$
\\
\multicolumn{1}{c}{   5.6E+12} &  &  &   &$(1.7 \pm 1.2) \times 10^{-6}$
 &$(1.9 \pm 1.1) \times 10^{-6}$  &$(7.4 \pm 2.1) \times 10^{-6}$
 &$(1.4 \pm 0.2) \times 10^{-5}$  &$(4.4 \pm 0.4) \times 10^{-5}$
 &$(4.1 \pm 0.4) \times 10^{-5}$  &$(7.7 \pm 1.8) \times 10^{-5}$
 &$(2.9 \pm 1.0) \times 10^{-5}$
\\
\multicolumn{1}{c}{   1.8E+13} &  &  &  &  &  & 
 &$(4.0 \pm 2.8) \times 10^{-7}$  &$(6.7 \pm 4.7) \times 10^{-7}$
 &$(1.7 \pm 0.6) \times 10^{-6}$  &$(6.4 \pm 1.3) \times 10^{-6}$
 &$(1.5 \pm 0.4) \times 10^{-6}$
\\
\hline
\end{tabular}
\end{scriptsize}
\end{sidewaystable}

\clearpage
\begin{sidewaystable}
\setlength{\tabcolsep}{1.25pt}
\begin{scriptsize}
\caption[Total IR Luminosity Function in the 100 $\mu$m band, OCGBH model]{Rest frame total IR 1/V$_{\ssty max}$ Luminosity Function in the PACS 100 $\mu$m band, assuming the OCGBH cosmological model. Units are dex$^{-1}$.Mpc$^{-3}$.
\label{vmtLIR100oGBH}}
\centering
\begin{tabular}{llllllllllll}
\hline \hline
\multicolumn{1}{c}{} &\multicolumn{ 11}{c}{Average redshift} \\
\multicolumn{1}{c}{Luminosity [L$_{\ssty \odot}$]}
 &\multicolumn{1}{c}{     0.2}  &\multicolumn{1}{c}{     0.4}
 &\multicolumn{1}{c}{     0.5}  &\multicolumn{1}{c}{     0.7}
 &\multicolumn{1}{c}{     0.9}  &\multicolumn{1}{c}{     1.1}
 &\multicolumn{1}{c}{     1.5}  &\multicolumn{1}{c}{     1.9}
 &\multicolumn{1}{c}{     2.2}  &\multicolumn{1}{c}{     2.8}
 &\multicolumn{1}{c}{     3.6}
\\
\hline
\multicolumn{1}{c}{   1.8E+08}  &$(4.7 \pm 3.9) \times 10^{-3}$ &  &  &  &  & 
&  &  &  &  & 
\\
\multicolumn{1}{c}{   5.6E+08}  &$(2.3 \pm 1.1) \times 10^{-3}$ &  &  &  &  & 
&  &  &  &  & 
\\
\multicolumn{1}{c}{   1.8E+09}  &$(6.7 \pm 1.7) \times 10^{-3}$ &  &  &  &  & 
&  &  &  &  & 
\\
\multicolumn{1}{c}{   5.6E+09}  &$(4.4 \pm 0.6) \times 10^{-3}$
 &$(7.1 \pm 5.3) \times 10^{-4}$ &  &  &  &  &  &  &  &  & 
\\
\multicolumn{1}{c}{   1.8E+10}  &$(2.6 \pm 0.2) \times 10^{-3}$
 &$(3.5 \pm 0.8) \times 10^{-3}$  &$(7.4 \pm 3.9) \times 10^{-4}$
 &$(9.2 \pm 9.2) \times 10^{-4}$ &  &   &$(7.3 \pm 7.3) \times 10^{-7}$ &  &  & 
& 
\\
\multicolumn{1}{c}{   5.6E+10}  &$(1.2 \pm 0.2) \times 10^{-3}$
 &$(2.3 \pm 0.2) \times 10^{-3}$  &$(2.6 \pm 0.4) \times 10^{-3}$
 &$(4.6 \pm 1.9) \times 10^{-3}$  &$(7.3 \pm 3.1) \times 10^{-4}$
 &$(9.1 \pm 9.1) \times 10^{-5}$ &  &  &  &  & 
\\
\multicolumn{1}{c}{   1.8E+11}  &$(1.9 \pm 0.3) \times 10^{-4}$
 &$(7.0 \pm 0.5) \times 10^{-4}$  &$(1.2 \pm 0.1) \times 10^{-3}$
 &$(2.2 \pm 0.2) \times 10^{-3}$  &$(1.5 \pm 0.2) \times 10^{-3}$
 &$(2.4 \pm 0.6) \times 10^{-3}$  &$(4.9 \pm 2.3) \times 10^{-4}$ &  &  &  & 
\\
\multicolumn{1}{c}{   5.6E+11}  &$(8.7 \pm 4.4) \times 10^{-6}$
 &$(4.7 \pm 1.1) \times 10^{-5}$  &$(1.3 \pm 0.2) \times 10^{-4}$
 &$(2.9 \pm 0.2) \times 10^{-4}$  &$(6.0 \pm 0.6) \times 10^{-4}$
 &$(6.5 \pm 1.1) \times 10^{-4}$  &$(7.2 \pm 1.0) \times 10^{-4}$
 &$(4.1 \pm 1.5) \times 10^{-4}$  &$(4.4 \pm 1.6) \times 10^{-4}$ &  & 
\\
\multicolumn{1}{c}{   1.8E+12}  &$(3.4 \pm 3.4) \times 10^{-6}$
 &$(3.0 \pm 3.0) \times 10^{-6}$  &$(3.3 \pm 2.3) \times 10^{-6}$
 &$(1.7 \pm 0.4) \times 10^{-5}$  &$(4.7 \pm 0.6) \times 10^{-5}$
 &$(1.03 \pm 0.10) \times 10^{-4}$  &$(1.5 \pm 0.1) \times 10^{-4}$
 &$(3.1 \pm 1.0) \times 10^{-4}$  &$(4.5 \pm 1.0) \times 10^{-4}$
 &$(8.8 \pm 5.2) \times 10^{-4}$  &$(1.9 \pm 1.4) \times 10^{-5}$
\\
\multicolumn{1}{c}{   5.6E+12} &  &  &   &$(9.4 \pm 9.4) \times 10^{-7}$
 &$(1.5 \pm 1.1) \times 10^{-6}$  &$(2.1 \pm 1.2) \times 10^{-6}$
 &$(9.3 \pm 1.6) \times 10^{-6}$  &$(4.4 \pm 1.4) \times 10^{-5}$
 &$(4.5 \pm 1.0) \times 10^{-5}$  &$(6.6 \pm 2.1) \times 10^{-5}$
 &$(2.7 \pm 1.7) \times 10^{-5}$
\\
\multicolumn{1}{c}{   1.8E+13} &  &  &  &  &  & 
 &$(2.3 \pm 2.3) \times 10^{-7}$  &$(8.1 \pm 5.7) \times 10^{-7}$
 &$(1.5 \pm 0.6) \times 10^{-6}$  &$(5.1 \pm 1.7) \times 10^{-6}$
 &$(2.9 \pm 1.1) \times 10^{-6}$
\\
\hline
\end{tabular}
\end{scriptsize}
\end{sidewaystable}

\begin{sidewaystable}
\setlength{\tabcolsep}{1.25pt}
\begin{scriptsize}
\caption[Total IR Luminosity Function in the 160 $\mu$m band, OCGBH model]{Rest frame total IR 1/V$_{\ssty max}$ Luminosity Function in the PACS 160 $\mu$m band, assuming the OCGBH cosmological model. Units are dex$^{-1}$.Mpc$^{-3}$.
\label{vmtLIR160oGBH}}
\centering
\begin{tabular}{llllllllllll}
\hline \hline
\multicolumn{1}{c}{} &\multicolumn{ 11}{c}{Average redshift} \\
\multicolumn{1}{c}{Luminosity [L$_{\ssty \odot}$]}
 &\multicolumn{1}{c}{     0.2}  &\multicolumn{1}{c}{     0.4}
 &\multicolumn{1}{c}{     0.5}  &\multicolumn{1}{c}{     0.7}
 &\multicolumn{1}{c}{     0.9}  &\multicolumn{1}{c}{     1.1}
 &\multicolumn{1}{c}{     1.5}  &\multicolumn{1}{c}{     1.9}
 &\multicolumn{1}{c}{     2.2}  &\multicolumn{1}{c}{     2.8}
 &\multicolumn{1}{c}{     3.6}
\\
\hline
\multicolumn{1}{c}{   1.8E+08}  &$(5.1 \pm 3.6) \times 10^{-3}$ &  &  &  &  & 
&  &  &  &  & 
\\
\multicolumn{1}{c}{   5.6E+08}  &$(6.2 \pm 3.0) \times 10^{-4}$ &  &  &  &  & 
&  &  &  &  & 
\\
\multicolumn{1}{c}{   1.8E+09}  &$(4.9 \pm 1.4) \times 10^{-3}$ &  &  &  &  & 
&  &  &  &  & 
\\
\multicolumn{1}{c}{   5.6E+09}  &$(5.8 \pm 1.3) \times 10^{-3}$
 &$(3.4 \pm 3.4) \times 10^{-4}$ &  &  &  &  &  &  &  &  & 
\\
\multicolumn{1}{c}{   1.8E+10}  &$(2.8 \pm 0.3) \times 10^{-3}$
 &$(3.9 \pm 1.0) \times 10^{-3}$  &$(6.7 \pm 3.1) \times 10^{-4}$
 &$(2.1 \pm 1.5) \times 10^{-4}$ &  &  &  &  &  &  & 
\\
\multicolumn{1}{c}{   5.6E+10}  &$(1.3 \pm 0.2) \times 10^{-3}$
 &$(1.9 \pm 0.2) \times 10^{-3}$  &$(3.1 \pm 0.5) \times 10^{-3}$
 &$(2.9 \pm 0.5) \times 10^{-3}$  &$(1.1 \pm 0.8) \times 10^{-4}$
 &$(7.6 \pm 7.6) \times 10^{-5}$ &  &  &  &  & 
\\
\multicolumn{1}{c}{   1.8E+11}  &$(2.1 \pm 0.3) \times 10^{-4}$
 &$(7.8 \pm 0.5) \times 10^{-4}$  &$(1.0 \pm 0.1) \times 10^{-3}$
 &$(1.4 \pm 0.2) \times 10^{-3}$  &$(1.3 \pm 0.2) \times 10^{-3}$
 &$(2.0 \pm 0.3) \times 10^{-3}$  &$(5.4 \pm 1.8) \times 10^{-4}$ &  &  &  & 
\\
\multicolumn{1}{c}{   5.6E+11}  &$(9.1 \pm 4.6) \times 10^{-6}$
 &$(5.1 \pm 1.2) \times 10^{-5}$  &$(1.3 \pm 0.2) \times 10^{-4}$
 &$(3.1 \pm 0.2) \times 10^{-4}$  &$(6.3 \pm 0.5) \times 10^{-4}$
 &$(6.4 \pm 0.9) \times 10^{-4}$  &$(6.3 \pm 0.8) \times 10^{-4}$
 &$(3.4 \pm 1.0) \times 10^{-4}$  &$(3.5 \pm 0.9) \times 10^{-4}$
 &$(4.8 \pm 3.4) \times 10^{-5}$ & 
\\
\multicolumn{1}{c}{   1.8E+12}  &$(3.5 \pm 3.5) \times 10^{-6}$
 &$(3.0 \pm 3.0) \times 10^{-6}$  &$(3.4 \pm 2.4) \times 10^{-6}$
 &$(1.8 \pm 0.4) \times 10^{-5}$  &$(4.8 \pm 0.6) \times 10^{-5}$
 &$(1.14 \pm 0.10) \times 10^{-4}$  &$(1.5 \pm 0.1) \times 10^{-4}$
 &$(1.7 \pm 0.3) \times 10^{-4}$  &$(3.4 \pm 0.6) \times 10^{-4}$
 &$(2.3 \pm 0.6) \times 10^{-4}$  &$(4.2 \pm 1.8) \times 10^{-5}$
\\
\multicolumn{1}{c}{   5.6E+12} &  &  &   &$(1.0 \pm 1.0) \times 10^{-6}$
 &$(1.7 \pm 1.2) \times 10^{-6}$  &$(2.2 \pm 1.3) \times 10^{-6}$
 &$(1.0 \pm 0.2) \times 10^{-5}$  &$(3.8 \pm 0.4) \times 10^{-5}$
 &$(4.1 \pm 0.4) \times 10^{-5}$  &$(6.1 \pm 1.4) \times 10^{-5}$
 &$(2.4 \pm 1.0) \times 10^{-5}$
\\
\multicolumn{1}{c}{   1.8E+13} &  &  &  &  &  & 
 &$(2.5 \pm 2.5) \times 10^{-7}$  &$(8.4 \pm 5.9) \times 10^{-7}$
 &$(1.5 \pm 0.6) \times 10^{-6}$  &$(6.0 \pm 1.3) \times 10^{-6}$
 &$(1.8 \pm 0.5) \times 10^{-6}$
\\
\hline
\end{tabular}
\end{scriptsize}
\end{sidewaystable}

\clearpage
\begin{table*} \tiny
\caption[Schechter parameters in the PACS 100 $\mu$m band.]{Best fitting Schechter parameters for the rest frame 100 $\mu$m 1/V$_{\ssty max}$ Luminosity functions.
\label{scht100}}
\begin{tabular}{lllllll}
\hline \hline
\multicolumn{1}{c}{} &\multicolumn{2}{c}{$\Lambda$CDM} &\multicolumn{2}{c}{CBGH}  &\multicolumn{2}{c}{OCBGH} \\
\multicolumn{1}{c}{$\bar{z}$}
 &\multicolumn{1}{c}{$\varphi^{\ssty \ast}$} &\multicolumn{1}{c}{L$^{\ssty \ast}$}
 &\multicolumn{1}{c}{$\varphi^{\ssty \ast}$} &\multicolumn{1}{c}{L$^{\ssty \ast}$}
 &\multicolumn{1}{c}{$\varphi^{\ssty \ast}$} &\multicolumn{1}{c}{$L^{\ssty \ast}$} \\ \hline
     0.2 &$(2.2 \pm 0.2) \times 10^{-3}$ &$(7.8 \pm 0.7) \times 10^{10}$
 &$(2.7 \pm 0.2) \times 10^{-3}$ &$(5.2 \pm 0.4) \times 10^{10}$
 &$(3.0 \pm 0.3) \times 10^{-3}$ &$(4.8 \pm 0.4) \times 10^{10}$\\
     0.6 &$(1.8 \pm 0.1) \times 10^{-3}$ &$(1.81 \pm 0.09) \times 10^{11}$
 &$(2.2 \pm 0.1) \times 10^{-3}$ &$(1.49 \pm 0.06) \times 10^{11}$
 &$(2.5 \pm 0.2) \times 10^{-3}$ &$(1.36 \pm 0.06) \times 10^{11}$\\
     1.0 &$(1.02 \pm 0.09) \times 10^{-3}$ &$(4.7 \pm 0.2) \times 10^{11}$
 &$(1.5 \pm 0.1) \times 10^{-3}$ &$(3.7 \pm 0.2) \times 10^{11}$
 &$(1.5 \pm 0.1) \times 10^{-3}$ &$(3.5 \pm 0.2) \times 10^{11}$\\
     1.5 &$(4.1 \pm 0.5) \times 10^{-4}$ &$(9.9 \pm 0.7) \times 10^{11}$
 &$(3.5 \pm 0.4) \times 10^{-4}$ &$(9.1 \pm 0.8) \times 10^{11}$
 &$(4.6 \pm 0.6) \times 10^{-4}$ &$(7.4 \pm 0.7) \times 10^{11}$\\
     2.1 &$(4.8 \pm 1.4) \times 10^{-4}$ &$(2.0 \pm 0.2) \times 10^{12}$
 &$(5.6 \pm 1.3) \times 10^{-4}$ &$(1.6 \pm 0.2) \times 10^{12}$
 &$(5.6 \pm 1.2) \times 10^{-4}$ &$(1.3 \pm 0.1) \times 10^{12}$\\
     3.0 &$(2.8 \pm 1.4) \times 10^{-4}$ &$(2.0 \pm 0.4) \times 10^{12}$
 &$(1.5 \pm 0.5) \times 10^{-4}$ &$(2.0 \pm 0.3) \times 10^{12}$
 &$(2.2 \pm 0.7) \times 10^{-4}$ &$(1.9 \pm 0.3) \times 10^{12}$\\
\hline
\end{tabular}
\end{table*}

\begin{table*} \tiny
\caption[Schechter parameters in the PACS 160 $\mu$m band.]{Best fitting Schechter parameters for the rest frame 160 $\mu$m 1/V$_{\ssty max}$ Luminosity functions.
\label{scht160}}
\begin{tabular}{lllllll}
\hline \hline
\multicolumn{1}{c}{} &\multicolumn{2}{c}{$\Lambda$CDM} &\multicolumn{2}{c}{CBGH}  &\multicolumn{2}{c}{OCBGH} \\
\multicolumn{1}{c}{$\bar{z}$}
 &\multicolumn{1}{c}{$\varphi^{\ssty \ast}$} &\multicolumn{1}{c}{L$^{\ssty \ast}$}
 &\multicolumn{1}{c}{$\varphi^{\ssty \ast}$} &\multicolumn{1}{c}{L$^{\ssty \ast}$}
 &\multicolumn{1}{c}{$\varphi^{\ssty \ast}$} &\multicolumn{1}{c}{$L^{\ssty \ast}$} \\ \hline
     0.2 &$(3.9 \pm 0.4) \times 10^{-3}$ &$(3.1 \pm 0.3) \times 10^{10}$
 &$(2.4 \pm 0.2) \times 10^{-3}$ &$(3.3 \pm 0.2) \times 10^{10}$
 &$(2.4 \pm 0.2) \times 10^{-3}$ &$(3.0 \pm 0.1) \times 10^{10}$\\
     0.6 &$(2.3 \pm 0.2) \times 10^{-3}$ &$(7.2 \pm 0.3) \times 10^{10}$
 &$(2.2 \pm 0.2) \times 10^{-3}$ &$(6.9 \pm 0.3) \times 10^{10}$
 &$(2.5 \pm 0.2) \times 10^{-3}$ &$(6.2 \pm 0.3) \times 10^{10}$\\
     1.0 &$(9.0 \pm 0.8) \times 10^{-4}$ &$(2.0 \pm 0.1) \times 10^{11}$
 &$(1.12 \pm 0.10) \times 10^{-3}$ &$(1.65 \pm 0.09) \times 10^{11}$
 &$(1.6 \pm 0.1) \times 10^{-3}$ &$(1.35 \pm 0.06) \times 10^{11}$\\
     1.5 &$(3.9 \pm 0.5) \times 10^{-4}$ &$(3.7 \pm 0.3) \times 10^{11}$
 &$(3.8 \pm 0.4) \times 10^{-4}$ &$(3.4 \pm 0.2) \times 10^{11}$
 &$(3.5 \pm 0.5) \times 10^{-4}$ &$(3.4 \pm 0.3) \times 10^{11}$\\
     2.1 &$(7.9 \pm 1.5) \times 10^{-5}$ &$(1.5 \pm 0.2) \times 10^{12}$
 &$(2.1 \pm 0.3) \times 10^{-4}$ &$(8.3 \pm 0.6) \times 10^{11}$
 &$(1.7 \pm 0.3) \times 10^{-4}$ &$(7.7 \pm 0.8) \times 10^{11}$\\
     3.0 &$(3.8 \pm 1.2) \times 10^{-5}$ &$(2.1 \pm 0.5) \times 10^{12}$
 &$(1.0 \pm 0.2) \times 10^{-4}$ &$(7.9 \pm 1.1) \times 10^{11}$
 &$(9.3 \pm 2.3) \times 10^{-5}$ &$(8.4 \pm 1.3) \times 10^{11}$\\
\hline
\end{tabular}
\end{table*}

\begin{table*} \tiny
\caption[Schechter parameters for the total IR, PACS 100 $\mu$m dataset.]{Best fitting Schechter parameters for the rest frame total IR 1/V$_{\ssty max}$ Luminosity functions
in the PACS 100 $\mu$m band. \label{schtLIR100}}
\begin{tabular}{lllllll}
\hline \hline
\multicolumn{1}{c}{} &\multicolumn{2}{c}{$\Lambda$CDM} &\multicolumn{2}{c}{CBGH}  &\multicolumn{2}{c}{OCBGH} \\
\multicolumn{1}{c}{$\bar{z}$}
 &\multicolumn{1}{c}{$\varphi^{\ssty \ast}$} &\multicolumn{1}{c}{L$^{\ssty \ast}$}
 &\multicolumn{1}{c}{$\varphi^{\ssty \ast}$} &\multicolumn{1}{c}{L$^{\ssty \ast}$}
 &\multicolumn{1}{c}{$\varphi^{\ssty \ast}$} &\multicolumn{1}{c}{$L^{\ssty \ast}$} \\ \hline
     0.2 &$(9.9 \pm 2.2) \times 10^{-4}$ &$(1.6 \pm 0.3) \times 10^{11}$
 &$(1.3 \pm 0.2) \times 10^{-3}$ &$(1.2 \pm 0.2) \times 10^{11}$
 &$(1.9 \pm 0.4) \times 10^{-3}$ &$(8.7 \pm 1.4) \times 10^{10}$\\
     0.4 &$(1.6 \pm 0.2) \times 10^{-3}$ &$(1.9 \pm 0.2) \times 10^{11}$
 &$(2.0 \pm 0.2) \times 10^{-3}$ &$(1.7 \pm 0.1) \times 10^{11}$
 &$(2.3 \pm 0.2) \times 10^{-3}$ &$(1.6 \pm 0.1) \times 10^{11}$\\
     0.5 &$(1.4 \pm 0.2) \times 10^{-3}$ &$(2.9 \pm 0.3) \times 10^{11}$
 &$(1.8 \pm 0.2) \times 10^{-3}$ &$(2.6 \pm 0.2) \times 10^{11}$
 &$(2.3 \pm 0.2) \times 10^{-3}$ &$(2.2 \pm 0.2) \times 10^{11}$\\
     0.7 &$(1.9 \pm 0.3) \times 10^{-3}$ &$(3.6 \pm 0.4) \times 10^{11}$
 &$(1.5 \pm 0.2) \times 10^{-3}$ &$(4.0 \pm 0.4) \times 10^{11}$
 &$(3.7 \pm 0.5) \times 10^{-3}$ &$(2.5 \pm 0.2) \times 10^{11}$\\
     0.9 &$(6.4 \pm 0.8) \times 10^{-4}$ &$(8.7 \pm 0.7) \times 10^{11}$
 &$(1.0 \pm 0.1) \times 10^{-3}$ &$(7.1 \pm 0.5) \times 10^{11}$
 &$(1.5 \pm 0.2) \times 10^{-3}$ &$(5.7 \pm 0.4) \times 10^{11}$\\
     1.1 &$(4.8 \pm 1.0) \times 10^{-4}$ &$(1.3 \pm 0.2) \times 10^{12}$
 &$(9.7 \pm 1.8) \times 10^{-4}$ &$(9.1 \pm 1.1) \times 10^{11}$
 &$(1.3 \pm 0.2) \times 10^{-3}$ &$(8.0 \pm 0.8) \times 10^{11}$\\
     1.5 &$(3.9 \pm 0.5) \times 10^{-4}$ &$(2.0 \pm 0.1) \times 10^{12}$
 &$(4.0 \pm 0.5) \times 10^{-4}$ &$(1.8 \pm 0.1) \times 10^{12}$
 &$(5.9 \pm 0.7) \times 10^{-4}$ &$(1.5 \pm 0.1) \times 10^{12}$\\
     1.9 &$(9.2 \pm 2.1) \times 10^{-5}$ &$(5.1 \pm 0.7) \times 10^{12}$
 &$(2.4 \pm 0.5) \times 10^{-4}$ &$(3.5 \pm 0.4) \times 10^{12}$
 &$(3.2 \pm 0.8) \times 10^{-4}$ &$(3.3 \pm 0.5) \times 10^{12}$\\
     2.2 &$(8.4 \pm 2.0) \times 10^{-5}$ &$(5.9 \pm 0.7) \times 10^{12}$
 &$(1.7 \pm 0.4) \times 10^{-4}$ &$(4.4 \pm 0.5) \times 10^{12}$
 &$(3.1 \pm 0.7) \times 10^{-4}$ &$(3.5 \pm 0.4) \times 10^{12}$\\
     2.8 &$(2.2 \pm 0.8) \times 10^{-4}$ &$(6.3 \pm 1.1) \times 10^{12}$
 &$(2.8 \pm 1.0) \times 10^{-4}$ &$(5.2 \pm 0.9) \times 10^{12}$
 &$(2.0 \pm 0.8) \times 10^{-4}$ &$(5.4 \pm 1.1) \times 10^{12}$\\
     3.6 &$(3.8 \pm 3.0) \times 10^{-6}$ &$(2.4 \pm 1.5) \times 10^{13}$
 &$(5.0 \pm 5.9) \times 10^{-6}$ &$(2.3 \pm 2.6) \times 10^{13}$
 &$(1.4 \pm 0.9) \times 10^{-5}$ &$(1.2 \pm 0.7) \times 10^{13}$\\
\hline
\end{tabular}
\end{table*}

\begin{table*} \tiny
\caption[Schechter parameters for the total IR, PACS 160 $\mu$m dataset.]{Best fitting Schechter parameters for the rest frame total IR 1/V$_{\ssty max}$ Luminosity functions
in the PACS 160 $\mu$m band. \label{schtLIR160}}
\begin{tabular}{lllllll}
\hline \hline
\multicolumn{1}{c}{} &\multicolumn{2}{c}{$\Lambda$CDM} &\multicolumn{2}{c}{CBGH}  &\multicolumn{2}{c}{OCBGH} \\
\multicolumn{1}{c}{$\bar{z}$}
 &\multicolumn{1}{c}{$\varphi^{\ssty \ast}$} &\multicolumn{1}{c}{L$^{\ssty \ast}$}
 &\multicolumn{1}{c}{$\varphi^{\ssty \ast}$} &\multicolumn{1}{c}{L$^{\ssty \ast}$}
 &\multicolumn{1}{c}{$\varphi^{\ssty \ast}$} &\multicolumn{1}{c}{$L^{\ssty \ast}$} \\ \hline
     0.2 &$(1.1 \pm 0.2) \times 10^{-3}$ &$(1.6 \pm 0.3) \times 10^{11}$
 &$(1.8 \pm 0.3) \times 10^{-3}$ &$(1.0 \pm 0.2) \times 10^{11}$
 &$(2.5 \pm 0.5) \times 10^{-3}$ &$(7.9 \pm 1.2) \times 10^{10}$\\
     0.4 &$(1.6 \pm 0.2) \times 10^{-3}$ &$(2.0 \pm 0.2) \times 10^{11}$
 &$(2.0 \pm 0.2) \times 10^{-3}$ &$(1.8 \pm 0.1) \times 10^{11}$
 &$(2.3 \pm 0.2) \times 10^{-3}$ &$(1.6 \pm 0.1) \times 10^{11}$\\
     0.5 &$(1.3 \pm 0.2) \times 10^{-3}$ &$(2.9 \pm 0.3) \times 10^{11}$
 &$(2.0 \pm 0.2) \times 10^{-3}$ &$(2.4 \pm 0.2) \times 10^{11}$
 &$(2.6 \pm 0.3) \times 10^{-3}$ &$(2.1 \pm 0.1) \times 10^{11}$\\
     0.7 &$(9.8 \pm 1.2) \times 10^{-4}$ &$(5.1 \pm 0.5) \times 10^{11}$
 &$(1.3 \pm 0.1) \times 10^{-3}$ &$(4.4 \pm 0.3) \times 10^{11}$
 &$(1.5 \pm 0.2) \times 10^{-3}$ &$(4.0 \pm 0.4) \times 10^{11}$\\
     0.9 &$(5.1 \pm 0.6) \times 10^{-4}$ &$(1.01 \pm 0.09) \times 10^{12}$
 &$(1.1 \pm 0.1) \times 10^{-3}$ &$(7.0 \pm 0.5) \times 10^{11}$
 &$(1.7 \pm 0.2) \times 10^{-3}$ &$(5.5 \pm 0.3) \times 10^{11}$\\
     1.1 &$(5.0 \pm 0.8) \times 10^{-4}$ &$(1.3 \pm 0.1) \times 10^{12}$
 &$(1.1 \pm 0.1) \times 10^{-3}$ &$(8.6 \pm 0.7) \times 10^{11}$
 &$(1.4 \pm 0.2) \times 10^{-3}$ &$(7.7 \pm 0.6) \times 10^{11}$\\
     1.5 &$(3.2 \pm 0.3) \times 10^{-4}$ &$(2.2 \pm 0.1) \times 10^{12}$
 &$(4.2 \pm 0.4) \times 10^{-4}$ &$(1.8 \pm 0.1) \times 10^{12}$
 &$(5.6 \pm 0.6) \times 10^{-4}$ &$(1.5 \pm 0.1) \times 10^{12}$\\
     1.9 &$(7.8 \pm 1.1) \times 10^{-5}$ &$(5.7 \pm 0.6) \times 10^{12}$
 &$(2.2 \pm 0.3) \times 10^{-4}$ &$(3.6 \pm 0.3) \times 10^{12}$
 &$(2.6 \pm 0.4) \times 10^{-4}$ &$(3.2 \pm 0.4) \times 10^{12}$\\
     2.2 &$(9.6 \pm 1.6) \times 10^{-5}$ &$(5.6 \pm 0.6) \times 10^{12}$
 &$(1.9 \pm 0.3) \times 10^{-4}$ &$(4.0 \pm 0.4) \times 10^{12}$
 &$(3.1 \pm 0.5) \times 10^{-4}$ &$(3.1 \pm 0.4) \times 10^{12}$\\
     2.8 &$(8.1 \pm 2.4) \times 10^{-5}$ &$(9.0 \pm 1.5) \times 10^{12}$
 &$(1.9 \pm 0.4) \times 10^{-4}$ &$(5.8 \pm 0.6) \times 10^{12}$
 &$(1.9 \pm 0.4) \times 10^{-4}$ &$(5.5 \pm 0.6) \times 10^{12}$\\
     3.6 &$(1.4 \pm 0.6) \times 10^{-5}$ &$(1.0 \pm 0.3) \times 10^{13}$
 &$(3.3 \pm 1.2) \times 10^{-5}$ &$(6.4 \pm 1.2) \times 10^{12}$
 &$(4.9 \pm 1.6) \times 10^{-5}$ &$(5.8 \pm 0.9) \times 10^{12}$\\
\hline
\end{tabular}
\end{table*}

\clearpage
\begin{table*}
\caption{Faint-end slopes values \label{alphat}}
\begin{tabular}{llll}
\hline \hline
dataset				&\multicolumn{1}{c}{$\Lambda$CDM}	&\multicolumn{1}{c}{CGBH}
&\multicolumn{1}{c}{OCGBH}		\\ \hline
$L_{\ssty 100 \mu m}$		&0.42 $\pm$ 0.04		&0.03 $\pm$ 0.05	&0.03 $\pm$ 0.05		\\
$L_{\ssty 160 \mu m}$		&0.25 $\pm$ 0.06		&0.00 $\pm$ 0.05	&0.00  $\pm$ 0.05				\\
$L_{\ssty IR, 100 \mu m}$	&0.67 $\pm$ 0.06		&0.38 $\pm$ 0.08	&0.33 $\pm$ 0.09		\\
$L_{\ssty IR, 160 \mu m}$	&0.61 $\pm$ 0.07		&0.26 $\pm$ 0.09	&0.2 $\pm$ 0.1			\\
\hline
\end{tabular}
\end{table*}

\begin{table}
{\tiny
\caption[Redshift evolution of the Schechter parameters]{Comoving number density and characteristic luminosity evolution parameters. \label{evolt}}
\centering
\begin{tabular}{llll}
\hline \hline
\multicolumn{1}{c}{dataset} &\multicolumn{1}{c}{model} &\multicolumn{1}{c}{$A$}		&\multicolumn{1}{c}{$B$} \\ \hline
 													&$\Lambda$CDM	&$(5.2 \pm 1.0) \times 10^{-1}$		&$(-3.3 \pm 0.6) \times 10^{-1}$ \\
$L_{\ssty 100 \mu m}$			&CGBH						&$(5.6 \pm 1.0) \times 10^{-1}$		&$(-4.5 \pm 0.8) \times 10^{-1}$ \\
 													&OCGBH					&$(5.5 \pm 0.9) \times 10^{-1}$		&$(-4.2 \pm 0.7) \times 10^{-1}$ \\
\multicolumn{4}{c}{} \\
 													&$\Lambda$CDM	&$(6.8 \pm 0.8) \times 10^{-1}$		&$(-7.6 \pm 0.6) \times 10^{-1}$ \\
$L_{\ssty 160 \mu m}$			&CGBH						&$(5.4 \pm 0.9) \times 10^{-1}$		&$(-5.8 \pm 0.8) \times 10^{-1}$ \\
 													&OCGBH					&$(5.4 \pm 0.9) \times 10^{-1}$		&$(-5.8 \pm 0.8) \times 10^{-1}$ \\
\multicolumn{4}{c}{} \\
 													&$\Lambda$CDM	&$(6.4 \pm 0.5) \times 10^{-1}$		&$(-6.6 \pm 1.0) \times 10^{-1}$ \\
$L_{\ssty IR, 100 \mu m}$	&CGBH						&$(6.4 \pm 0.4) \times 10^{-1}$		&$(-6.3 \pm 0.9) \times 10^{-1}$ \\
 													&OCGBH					&$(6.3 \pm 0.5) \times 10^{-1}$		&$(-6.1 \pm 0.7) \times 10^{-1}$ \\
\multicolumn{4}{c}{} \\
 													&$\Lambda$CDM	&$(5.8 \pm 0.7) \times 10^{-1}$		&$(-5.9 \pm 0.5) \times 10^{-1}$ \\
$L_{\ssty IR, 160 \mu m}$	&CGBH						&$(5.6 \pm 0.7) \times 10^{-1}$		&$(-5.1 \pm 0.3) \times 10^{-1}$ \\
 													&OCGBH					&$(5.6 \pm 0.7) \times 10^{-1}$		&$(-5.1 \pm 0.3) \times 10^{-1}$ \\ \hline
\end{tabular}}
\end{table}

\bibliographystyle{aa}
\bibliography{general.bib}

\appendix
\section{Mock Catalogues}
\label{mock}
In this appendix we test whether the 1/V$_{\ssty max}$ LF estimator is reliable
in studying Gpc scale voids like the ones proposed by the GBH models, embedded
in an LTB dust model. We follow the general approach by \citep{2000ApJS..129....1T} who
made use of mock catalogues that were built assuming a (non-central and small) void with
a radius of 1.6 Mpc at a distance of 0.8 Mpc, and at a limiting redshift of $z = 0.1$.
Our mock catalogues are built using the matter density distributions in the GBH models,
as shown in Figure \ref{plotom}. Also, the redshift range of our interest
is 4 times larger, since we want to test the validity of the estimator in the interval
$\Delta z = [0.01,0.4]$, where we fit the faint-end slope of the luminosity functions.

Mock catalogues are built reproducing the detection limits and
SED distributions in the GOODS-S and COSMOS fields, in the PACS 100 and 160 $\mu$m filters,
as listed in \citep{2013MNRAS.tmp.1158G}. We chose those two fields for better representing the whole of
the data used in this work: GOODS-S is the field with the lowest flux
limits in the PEP survey, while COSMOS is the one with the widest effective area.

Naively one might decide to use matter density distributions in Figure \ref{plotom} to
randomly assign comoving distances to the sources in the mock catalogue. However, given the
large redshift interval we aim to cover in our simulations, the redshift evolution of the
density profiles must be fully considered.

For each of the present time, rest-frame (z=0) matter density profile, $\Omega_{\ssty M}(r)$,
defined in a constant time coordinate hipersurface, and fit by both the standard model and
the void models (see Figure \ref{plotom}), we compute the corresponding redshift evolution,
$\Omega_{\ssty M}(z)$, defined in the past light-cone of the same cosmological model. In the
FLRW spacetime, the dimensionless density parameters $\Omega_{\ssty M}$ and 
$\Omega_{\ssty M}(z)$ are related as follows:
\begin{equation}
\label{FLRWomz}
\Omega_{\ssty M}(z) = \Omega_{\ssty M} \left[ \frac{H_{\ssty 0}}{H(z)}\right]^2 a(z)^{-3},
\end{equation}
where $H(z)$ is the Hubble parameter at redshift $z$, carried over from the definition of the critical
density $\rho_{\ssty c}=3H_{\ssty 0}^2/8\pi G$, and $a(z)$ the scale factor, both as functions of the
redshift. Similarly, following the definition of $\Omega_{\ssty M}(r)$ used in GBH and
\citep{2012JCAP...10..009Z}, one may write an analogue equation in the void-LTB models as,
\begin{equation}
\label{LTBomz}
\Omega_{\ssty M}(z) = \Omega_{\ssty M}[r(z)] \left\{ \frac{H_{\ssty \perp 0}[r(z)]}
{H_{\ssty \perp}[t(z),r(z)]}\right\}^2 a_{\ssty \perp}[t(z),r(z)]^{-3},
\end{equation}
where $H_{\ssty \perp}[t(z),r(z)]$ and $a_{\ssty \perp}[t(z),r(z)]$ are now the transverse
Hubble parameter and scale factor, respectively. Figure \ref{plotomz} shows the redshift
evolution of the density parameters in the three models considered in the present work.
Note, however, that there's an ambiguity in the definition of equation (\ref{LTBomz}),
due to the fact that the LTB geometry possesses radial expansion rate and scale factor
that are in general different from their transverse counterparts. For the purpose of
building mock catalogues that are consistent with the void-LTB parametrizations used in
this work, we chose to use the transverse quantities, because those were the ones
used in \citep{2012JCAP...10..009Z}, from where the best fit parameters used in this work
were taken.

Next, we randomly assign: a. redshifts using a probability distribution based on
one of those $\Omega_{\ssty M}(z)$ profiles; b. rest-frame luminosities, based
on an input Schechter LF with parameters L$^*$=10$^{11}$ L$_{\ssty \sun}$, $\phi^*$=10$^{-3}$
dex$^{-1}$.Mpc$^{-3}$ and $\alpha$ = -1/2; and c. a representative empirical SED from the Poletta
templates, drawn from the same distributions reported in \citep{2013MNRAS.tmp.1158G}. In this way, we can test first
the validity of the 1/V$_{\ssty max}$ estimator itself for the purposes of the present work,
and second, the possible effects of the different predicted density profiles on the values
recovered of the LF.

Having assigned a redshift, a luminosity and a SED for each Monte Carlo (MC) realisation, we proceed
to compute k-corrections and fluxes, using the luminosity distance-redshift relation consistent
with the cosmology assumed for the redshift assignment. We include the source in the mock catalogue
if its observed flux is larger than the detection limit of the field. We repeat such
process until we have a catalogue with a number of selected MC realisations equal to the number
of sources in the redshift interval $\Delta z = [0.01,0.4]$, for a given field.

\begin{figure}
\centering
\includegraphics[width=8.5cm]{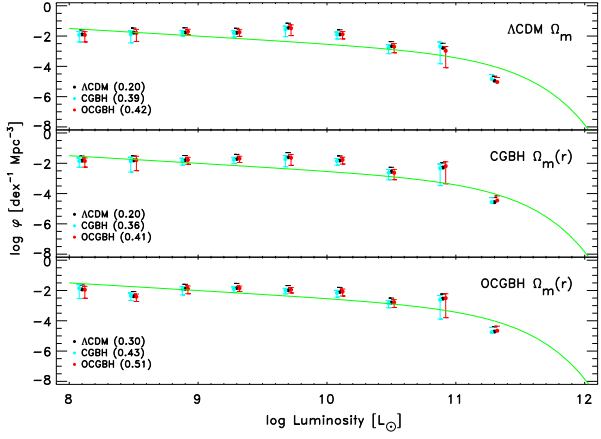}
\caption{Results for the 1/V$_{\ssty max}$ LF estimator, computed from mock catalogues
assuming a constant density profile $\Omega_{\ssty M}=0.27$ ($\Lambda$CDM), and the underdense
profiles of equation (\ref{GBHOM}) for the GBH void models (Figure \ref{plotom}). Sources luminosities
 in the mock catalogue are drawn from the Schechter LF (here shown in green dashed line,
with parameters L$^*$=10$^{11}$ L$_{\ssty \sun}$, $\varphi^*$=10$^{-3}$ dex$^{-1}$.Mpc$^{-3}$ and
$\alpha$ = -1/2). Flux limits and SED are taken from the results of
\citep{2013MNRAS.tmp.1158G} for the PEP survey dataset in the GOODS-S field. \label{plotmock_rgs}}
\end{figure}

\begin{figure}
\centering
\includegraphics[width=8.5cm]{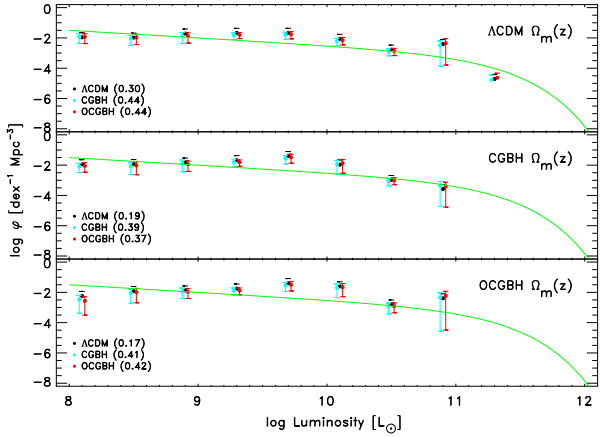}
\caption{Same as Figure~\ref{plotmock_rgs} but assuming the redshift evolution of the matter density profiles in both the standard ($\Lambda$CDM),
and the void-LTB models as in Figure \ref{plotomz}.\label{plotmock_zgs}}
\end{figure}

\begin{figure}
\centering
\includegraphics[width=8.5cm]{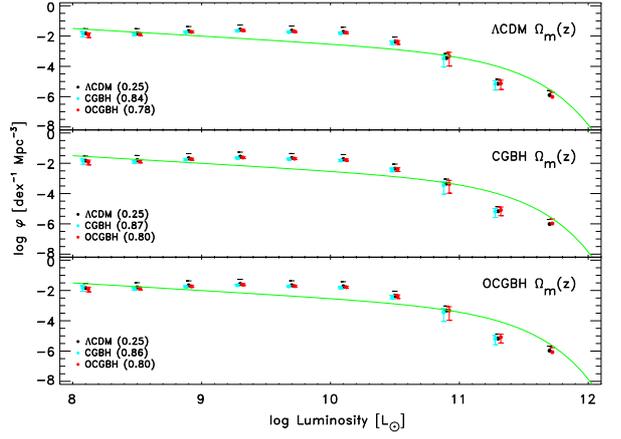}
\caption{Same as Figure~\ref{plotmock_zgs} but for the PEP survey dataset in the COSMOS field.\label{plotmock_zcm}}
\end{figure}

We then compute the 1/V$_{\ssty max}$ LF following the same methodology described in
\S \ref{vmaxlfsec}, using 100 mock catalogues built as above. To assess the goodness-of-fit of
the 1/V$_{\ssty max}$ LF versus the input Schechter profile, we compute the one-sided
Kolmogorov-Smirnov (KS) statistic of the normalised residuals against a Gaussian with zero mean
and unit variance. We plot the 1/V$_{\ssty max}$ points computed using the mock catalogues against
the input Schechter LF used in their build-up in Figures \ref{plotmock_rgs}, \ref{plotmock_zgs},
and \ref{plotmock_zcm}
The KS statistic for each mock/input comparison is listed between parentheses,
in the plots. The smaller this value, the closer the normalized residuals are to a gaussian
with zero mean and unit variance.

We find that 
the matter density parameter profiles of interest don't change
significantly the LF results, as can be seen by comparing different panels in a same Figure.
We note what appears to be a general bias towards under-estimating the characteristic luminosity
$L^\ast$, in agreement with \citep{2012MNRAS.426..531S} results.

Comparison between the $1/V_{\ssty max}$ LF results for the GOODS-S mocks built using either
the present time density profiles (Figure \ref{plotmock_rgs}) or the appropriate redshift evolution
of those (Figure \ref{plotmock_zgs}), shows that the method successfully takes into consideration
the redshift distortion in the matter distribution, yielding points in both cases that recover the
input LF profile qualitatively close in respect to each other, with respect to their KS
statistics. 

Comparison between the mocks for the GOODS-S (Figure \ref{plotmock_zgs}) and COSMOS
(Figure \ref{plotmock_zcm}) fields built using the redshift evolution of the density parameter
in the different cosmological models shows that the $1/V_{\ssty max}$ estimator
fares slightly better in the deeper GOODS-S field, as compared to the wider COSMOS one.

Summing up, even if the method is not perfectly robust under a change in the cosmological model,
the variations caused by a change in the underlying cosmology in the results obtained with the
$1/V_{\ssty max}$ estimator are not enough to explain the significant differences in the shape
of the LF at the considered redshift interval, $\Delta z = [0.01,0.4]$.


\end{document}